\documentclass[preprint,10pt,3p]{elsarticle}

\usepackage{amsmath,amssymb,amsfonts}
\usepackage{algorithmic}
\usepackage{graphicx}
\usepackage{textcomp}
\usepackage{url}
\usepackage{fontawesome5}
\usepackage[T1]{fontenc}
\usepackage[hidelinks]{hyperref}
\usepackage[most,breakable]{tcolorbox}
\usepackage{longtable,capt-of}
\usepackage{array}

\journal{Journal of Systems and Software}

\begin{document}

\begin{frontmatter}



\title{Exploring Ethical Concerns of Mobile Applications from App Reviews: A Literature Survey}


\author[1]{Aakash Sorathiya}
\ead{aakash.sorathiya@ucalgary.ca}

\author[1]{Gouri Ginde\corref{cor1}}
\ead{gouri.ginde@ucalgary.ca}

\cortext[cor1]{Corresponding author}
\affiliation[1]{organization={Department of Electrical and Software Engineering, University of Calgary},
            city={Calgary},
            country={Canada}}

\begin{abstract}
Privacy, security, and accessibility, like ethical concerns in mobile applications (a.k.a. apps), commonly subsumed under non-functional requirements, are generally reported by users through app reviews available in app stores. However, these remain unidentified among other types of reviews, such as user experiences, problem reports, and new feature discussions. Over the past decade, extensive research has focused on extracting valuable information from app reviews, including feature requests and bug reports. However, there remains a lack of a synthesis of research related to app review analysis for exploring users' ethical concerns. This paper presents a comprehensive survey of this research area, covering 37 relevant studies published since 2012, identified from the initial 553 studies using specific inclusion and exclusion criteria. The studies examined vary in review counts, ranging from 500 to 626 million, and include between a single and 1.3 million apps. Our detailed analysis highlights diverse objectives, methodologies, and strategies, along with additional resources such as app privacy policies, which researchers generally utilize to analyze ethical concerns. Our findings also identify persistent barriers to privacy, security, accessibility, transparency, fairness, accountability, and safety, as reported by users in app reviews. Furthermore, we propose a research agenda that focuses on four key areas, including automated extraction and classification of ethical concerns-related app reviews. Our survey outcomes can assist developers and system architects in recognizing and prioritizing non-functional requirements at the initial stages of the development lifecycle, whereas researchers can expand upon this synthesis to create tools for the automated detection of ethical concerns.

\end{abstract}




\end{frontmatter}



\section{Introduction}
Mobile applications (mobile apps) are meticulously developed taking into account particular user objectives, such as enhancing social capital in Sharing Economy apps \cite{martin2016sharing} or fostering healthy practices in Health\&Fitness apps \cite{dennison2013opportunities}. However, due to intense market rivalry, the app development cycle often aims to produce functional apps within short time frames (such as days or weeks), leading developers to deviate from their initial objectives frequently. These frequent divergences raise ethical concerns, such as privacy, security, safety, or transparency \cite{gillespie2019you, hill2022wrongfully, zuboff2019age}, which are also considered software quality attributes or non-functional requirements \cite{chung2012non, chung1993dealing, breaux2008analyzing, manasreh2025designing, liywalii2024requirements, martinez2022software}. Mobile apps that do not adequately implement these requirements are often labeled as untrustworthy or even abandoned by their users \cite{anderson2016mobile}. Thus, for apps to endure market scrutiny, developers continuously keep track of user feedback through a valuable resource: app reviews available on the app stores \cite{palomba2018crowdsourcing}, which has proved to be a useful datasource for many areas of software engineering (SE) (e.g., requirement engineering, testing, etc.) \cite{biswas2024interpretable}.

Numerous studies provide compelling evidence that developers utilize app reviews to strategically plan for upcoming releases, thereby enhancing both ratings and overall reputation. They typically analyze user feedback to gain valuable insights regarding bug reports, feature requests, network connectivity issues, resource consumption challenges (including battery performance), and user interface difficulties \cite{palomba2015user, palomba2018crowdsourcing, li2018mobile}. Significant progress has been achieved in this domain, encompassing the creation of tools and specialized methodologies designed to translate app reviews into actionable requirements \cite{chen2014ar, pelloni2018becloma}.


Recently, a considerable number of researchers have been meticulously conducting a thorough analysis of app reviews with the primary objective of delving deeply into users' ethical concerns, such as privacy, security, and accessibility, in relation to mobile apps. The overarching goal of these researchers is to find compelling evidence from user feedback that reveals the significant ethical concerns posed by mobile apps, thus assisting app developers in proactively addressing these pressing concerns during the initial phases of the app development process. However, despite the efforts in this area, the current landscape of research and advancements in the analysis of app reviews aimed at exploring ethical concerns, as well as the practical application of this acquired knowledge, is still limited, particularly since there has not yet been a literature survey conducted within this specific domain. This highlights the pressing need for a comprehensive survey of the related literature and a systematic mapping of existing studies to effectively bridge the existing knowledge gap by thoroughly analyzing the role and impact of app reviews in understanding and mitigating users' ethical concerns. Consequently, this paper seeks to contribute to enhancing the understanding of this critical domain.

Our primary objective in this survey is to deliver a comprehensive analysis of the role that app reviews play in recognizing users' ethical concerns regarding mobile apps, while concurrently exploring the associated challenges, methodologies, and potential avenues for future research.

The primary contributions of our literature survey are:
\begin{itemize}
    \item \textit{Enhanced knowledge and practical implications}: In this literature survey, we provide researchers and practitioners with a deeper understanding of the current landscape of research related to the app review analysis for exploring users' ethical concerns. This includes valuable insights into diverse methodologies, significant findings, and trends uncovered through meticulous systematic analysis. Our findings can also assist development teams in tackling ethical concerns during the requirements engineering phase, mitigating risks, and ensuring adherence to legal standards. The outcome of this research can benefit developers, policymakers, and researchers. For example, these findings can be further operationalized in requirements engineering practices, automated review mining tools, or regulatory compliance.
    \item \textit{Publication trends and knowledge gaps}: We highlight the emergence of significant trends while concurrently identifying gaps in the current literature related to app review analysis and ethical concerns. Through the synthesis of findings from various studies, our research highlights the areas that necessitate additional exploration or advancement. For instance, we found that many studies focus heavily on privacy and security concerns, while other ethical aspects, such as fairness, accountability, and safety, receive much less attention. This imbalance highlights a significant gap in the field, as it shows that researchers focus on the most technically measurable ethical concerns while overlooking others equally critical to user trust and inclusion.
    \item \textit{Future research directions}: Also, we highlight detailed future work challenges and provide strategic insights for forthcoming research directions by revealing prospective avenues for the advancement of this field.
\end{itemize}

This paper is organized as follows. Section \ref{rw} presents the background and related work. Section \ref{design} outlines the research methodology, detailing research questions (RQs), search strategy, research process, selection criteria, data extraction strategy, and data synthesis. Section \ref{results} presents the analysis of the results, addressing our RQs, and Section \ref{discuss} presents the implications for software engineers. Section \ref{threats} examines the threats to validity, and finally, Section \ref{conclude} offers concluding remarks and suggests future research directions.

\section{Background and Related Work} \label{rw}
\subsection{Ethical Concerns} \label{ec}
As digital technologies, specifically mobile apps and artificial intelligence (AI) systems, become increasingly embedded in critical domains, ranging from healthcare and finance to governance and surveillance, they raise significant ethical concerns that must be addressed systematically. The moral principles and professional standards that must be considered during software design, development, deployment, and use are encompassed as ethical concerns of software \cite{wright2011framework}. Ensuring ethical integrity is fundamental to complying with regulatory frameworks and sustaining public trust, preventing harm, and promoting equitable technological advancement \cite{wright2011framework}. Ethical concerns often overlap with broader concepts such as human-centric issues and socio-technical concerns; however, we adopt a more focused definition in our study. Human-centric issues generally relate to usability, user experience, and user satisfaction, as commonly discussed in human-centered design literature \cite{chung2012non, chung1993dealing}. Socio-technical concerns emphasize the interplay between social contexts and technical systems, including organizational, cultural, and institutional factors \cite{wright2011framework}.

Although many ethical issues have been discussed in the literature, this paper focuses on seven main concerns: privacy, security, accessibility, transparency, fairness, accountability, and safety. These seven concerns were selected not only because they frequently appear in prior studies, but also because they are grounded in established ethical frameworks and guidelines for software and AI systems \cite{wright2011framework, kaur2022trustworthy, european-commission-2019}. In particular, prior work in software ethics and responsible AI consistently highlights principles such as privacy, fairness, transparency, accountability, and safety as core dimensions of ethical system design, while accessibility is widely recognized as essential for inclusive and equitable technology use. Together, these concerns represent key ethical areas that are widely accepted in both academic and policy discussions. In the following, we briefly describe these concerns using the definitions from existing studies \cite{wright2011framework, kaur2022trustworthy, european-commission-2019}.


\textit{Privacy} concern refers to the unauthorized use, access, or disclosure of user's personal and confidential data.

\textit{Security} concern refers to malicious attacks due to software vulnerabilities, data corruption, or destruction or modification of personal data.

\textit{Accessibility} concern refers to the lack of inclusivity and usability of the software affecting user groups based on their age, gender, disability, or other socio-economic attributes.

\textit{Transparency} concern refers to the inability of software to provide a clear understanding of how it works and on what basis it makes decisions.

\textit{Fairness} concern refers to the unfair decision-making by software that is biased or prejudiced toward any user or user group.

\textit{Accountability} concern refers to the inability of a software or software company in taking responsibility for the decisions and actions taken by the software. It also refers to the inability of a software company to identify and address the negative and positive consequences of the software and taking responsibility for it.

\textit{Safety} concerns refer to the physical or psychological harm to users when using software.

Developers and policymakers can better ensure that technologies conform to legal requirements, human rights, and societal values by tackling these fundamental ethical concerns. We used this set of common ethical concerns further in our survey to identify relevant studies and categorize them into these concerns.

\subsection{Related Work}
To the best of our understanding, there appears to be no comprehensive survey that integrates the app review analysis with the analysis of ethical concerns for mobile apps within the existing body of literature. However, independent surveys have offered a comprehensive synthesis of these methodologies.

For instance, Genc-Nayebi and Abran \cite{genc2017systematic} conducted a systematic literature review (SLR) analyzing 24 primary studies published between 2000 and 2015, focusing on mobile app store data mining, opinion aggregation, and spam detection. Their analysis revealed a growing attraction in app review mining and identified significant challenges, including the unstructured nature of app reviews, the colloquial language used by users, the domain and context dependence of app reviews, and the identification of spam or fake reviews. For future research, the study suggested the need for broader and deeper commonsense knowledge bases for app reviews mining systems and integration of these systems into every app or mobile device. Another review from D{\k{a}}browski et al. \cite{dkabrowski2022analysing} analyzed 182 papers published between 2012 and 2020, focusing on app reviews analysis for SE activities. Their analysis revealed a growing trend in app reviews analysis to support various SE tasks, such as requirements elicitation, requirements prioritization, and requested modification prioritization. The study also identified the need for a greater focus on SE goals and use cases to increase the relevance and impact of app review analysis techniques.

Several surveys have analyzed singular ethical concerns for mobile apps. For example, Alshawi et al. \cite{alshawi2022data} conducted an SLR analyzing 40 primary studies to evaluate the data privacy of COVID-19 mobile apps. Their analysis identified privacy issues related to data storage, retention, access, monitoring, integrity, compliance, and disclosure, and suggested solutions to improve data privacy from three perspectives: public, law, and health considerations. Masruroh et al. \cite{masruroh2022evaluation} addressed the evaluation of usability and accessibility of mobile apps for people with disabilities by analyzing 15 primary studies in an SLR. Their analysis highlighted seven focus evaluation areas, including accessibility, and identified fourteen evaluation methods; however, no standardized method was universally accepted by usability evaluators, leading to the combined use of various methods to achieve optimal evaluations of mobile apps.

Our study extends these works in three important ways.
First, we integrate the perspective of ethical concerns, including privacy, security, accessibility, fairness, accountability, transparency, and safety, into the analysis of app review research, providing a thematic synthesis that has not been attempted before.
Second, unlike previous surveys that primarily summarized data mining techniques, we critically evaluate how ethical issues are identified, classified, and validated, revealing methodological patterns and weaknesses across studies.
Third, we map underexplored ethical categories and contradictions in findings to propose a focused research agenda for this emerging intersection of ethics and SE.
In doing so, this survey moves beyond simply cataloguing existing approaches. It provides a comprehensive overview of how app review analysis contributes to understanding users’ ethical concerns in mobile applications.
We formulated a set of four research questions to guide this literature survey process and conducted a comprehensive search to identify publications that address these research questions. We also delved deeper into this domain and recognized numerous potential avenues for future investigation with four key research directions, such as automated extraction and classification of ethical concerns-related app reviews, and conducting interviews with focus groups to understand direct real-world perspectives.


\section{Research Methodology} \label{design}
\begin{figure*}[h]
    \centering
    \includegraphics[width=1\linewidth]{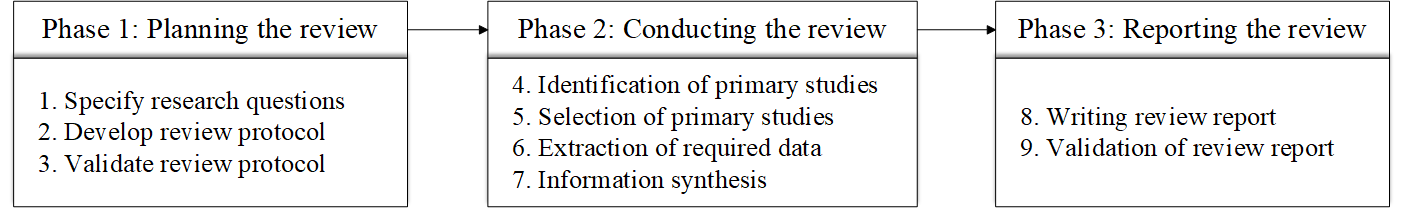}
    \caption{Process of our literature survey based on the guidelines outlined by Kitchenham \cite{kitchenham2004procedures} to conduct SLRs, which has three phases to it.}
    \label{fig:slr_process}
\end{figure*}

In this section, we discuss the process of our literature survey, based on the guidelines of Kitchenham \cite{kitchenham2004procedures} to conduct systematic literature reviews, which involved three main phases: planning, conducting, and reporting the review as shown in Figure~\ref{fig:slr_process}. Each phase included specific tasks outlined in the sections below.

\subsection{Planning}
The planning phase clarified the precise goals of the SLR, that is, to explore how app reviews contribute to identifying and addressing users' ethical concerns. It encompassed formulating research questions, selecting keywords, conducting pilot searches, exploring scientific databases, and defining selection criteria and data extraction fields. We have outlined the main components of the planning phase below.

\subsubsection{Research questions}
\textit{RQ1: What are the attributes of the collected app reviews, and what techniques were used in their collection?} This question aims to determine the number of apps and reviews analyzed in each paper, the terminology and language of app reviews, and whether they focus on specific app domains. Additionally, it seeks to understand the data collection methods, such as manual collection, proprietary platforms, or automated scripts.

\noindent\textit{RQ2: What objectives, methods, and resources are frequently utilized to analyze app reviews related to ethical concerns?} This question explores the objectives of analyzing app reviews, as well as the analysis stages and research methodologies employed by the studies for app review analysis, including the prevalence of quantitative or qualitative methods, and the additional resources and tools adopted.

\noindent\textit{RQ3: What particular ethical concerns-related issues do users report in the analyzed app reviews?} This question addresses the persistent barriers reported by users in the app reviews related to ethical concerns.

\noindent\textit{RQ4: What challenges for future research have been identified?} This question refers to future research directions highlighted by the primary studies. It intends to explore past and present research, identifying potential avenues for exploring and addressing users' ethical concerns through app reviews. Additionally, it seeks to provide a research agenda to guide future researchers and foster significant advancements in this area.

\subsubsection{Development and validation of the review protocol}
Our review protocol outlines the necessary steps that were employed to conduct this comprehensive literature survey. 

\textbf{Search scope. }The following four electronic databases were searched for primary studies according to Kitchenham's  \cite{kitchenham2004procedures} suggestions. We selected Scopus and Web of Science, as they index many publishers, including the most common publishers of computer science and software engineering research (e.g. Elsevier, Springer, etc.). Furthermore, we included two databases of venue-specific publishers, IEEE and ACM, since they were identified as the most common publishers for research on the analysis of app reviews in the SLR of D{\k{a}}browski et al. \cite{dkabrowski2022analysing}.
\begin{itemize}
    \item Scopus (\url{www.scopus.com})
    \item Web of Science (WoS) (\url{www.webofscience.com})
    \item ACM Digital Library (DL) (\url{https://dl.acm.org})
    \item IEEE Xplore (\url{https://xplorestaging.ieee.org})
\end{itemize}

\textbf{Search terms. }Following the PICO (\textbf{P}opulation, \textbf{I}ntervention, \textbf{C}omparison, and \textbf{O}utcome) criteria from Kitchenham's  \cite{kitchenham2004procedures} guidelines, where the population of this study refers to ``Ethical concerns'' and ``App reviews'', we developed the initial search query as \textit{(``app reviews'' AND ``ethical concerns'')}. We only considered population as search terms in this study, because we are not analyzing any particular intervention or comparing it with other interventions in app reviews mining.

Using this noncomprehensive search string, we conducted the pilot search in the selected electronic databases and identified the synonyms of ``app reviews'' used in various studies: ``user reviews'', ``user feedback'', ``user comments'', and ``user perspectives''. Additionally, we identified that most of the studies did not use the broader term ``ethical concerns'', instead, they referred to the specific ethical concern that they analyzed, like privacy or accessibility. Therefore, to create a comprehensive list of specific ethical concerns to be added in our search string, we referred to these seven commonly occurring ethical concerns: privacy, security, accessibility, accountability, transparency, fairness, and safety, which we discussed in Section \ref{rw}. There are many possible ethical issues in software, but it is not realistic to include all of them in a single review because the terminology and focus vary greatly between studies. As outlined in Section \ref{ec}, these seven concerns are grounded in established ethical frameworks and represent the dimensions most consistently addressed in prior work, providing a focused yet comprehensive scope for this review.

Additionally, we included ``mobile applications'' in our search string to focus exclusively on studies in the domain of mobile apps. Further, we used the regex pattern in the search query to encompass the main term and its variants. For example, we used ``app*'' to encompass ``application'' and its variations such as ``apps'' or ``applications''. We also opted to search within the abstracts, as they provide a larger sampling of texts, in addition to titles and keywords. Finally, we developed the following comprehensive search query.

\begin{tcolorbox}[colback=gray!10!white,colframe=gray!75!black]
( ``user reviews'' OR ``user feedback'' OR ``user comments'' OR ``user perspectives'' OR ``app reviews'' ) AND ( ethical* OR priva* OR secur* OR transpare* OR accessi* OR accounta* OR fair* OR safe* ) AND ( ``mobile app*'' )
\end{tcolorbox}

\textbf{Study selection criteria. }After the search terms were set and databases were selected, we defined the inclusion and exclusion criteria (shown in Table \ref{tab:icec}) to select the relevant studies from the search results accordingly. We defined eight exclusion criteria (EC1 to EC8) and two inclusion criteria (IC1 and IC2). 
We excluded studies that met any of the exclusion criteria and included only those studies that satisfied both inclusion criteria..

\begin{table}[h]
    \centering
    \caption{Inclusion and exclusion criteria.}
    \label{tab:icec}
    \begin{tabular}{p{15cm}}
        \textbf{Inclusion criteria (IC)} \\
        \hline
        \\
        \textbf{IC1}. Papers that specifically analyzed app reviews as a method for assessing users' ethical concerns. \\
        \textbf{IC2}. Papers that provided empirical findings, insights, or recommendations related to mobile app privacy, security, accessibility, accountability, fairness, transparency, and safety. \\
        \hline
        \\
        \textbf{Exclusion criteria (EC)} \\
        \hline
        \\
        \textbf{EC1}. Papers published before 2012. \\
        \textbf{EC2}. Papers not written in English. \\
        \textbf{EC3}. Papers unavailable in full-text or duplicated. \\
        \textbf{EC4}. Papers that were conference abstracts, posters, or vision papers without sufficient detail. \\
        \textbf{EC5}. Non-peer-reviewed papers such as those published in magazines, notes, etc. \\
        \textbf{EC6}. Pre-expansion conference paper of the extended journal or conference paper. \\
        \textbf{EC7}. Papers that analyzed app reviews but did not mention any ethical concerns. \\
        \textbf{EC8}. Papers that analyzed mobile app ethical concerns but did not consider app reviews as their data source.\\
        \hline
    \end{tabular}
\end{table}

\textbf{Study selection procedure. }We systematically divided the study selection procedure into three rounds. In Round 1, we read only the title and abstract of each paper and used EC1-EC6 exclusion criteria to decide whether to exclude it. We moved all the remaining papers to the next round. In Round 2, we read both the introduction and conclusion of all the remaining papers. Same as before, we retained these papers until the next round, except for the excluded papers. In Round 3, we read the full text, and then the final decision was made to select the study for our survey. In Rounds 2 and 3, we utilized EC7 and EC8 to exclude the studies and IC1 and IC2 to include the studies.

\textbf{Snowball search. }To make our study more comprehensive, we conducted a one-step backward and forward snowball search \cite{wohlin2014guidelines} by scanning the references of the papers selected via the previous procedure. We used Sematic Scholar APIs\footnote{\url{https://api.semanticscholar.org/graph/v1/paper/{paper_id}/citations}}\footnote{\url{https://api.semanticscholar.org/graph/v1/paper/{paper_id}/references}} to automatically fetch the references and citations of the selected studies. After fetching all the papers, we filtered the irrelevant papers by applying our search strings and followed the selection procedure described above.

\subsection{Conducting the review}

\subsubsection{Identification and selection of the primary studies}
\begin{figure}[htb]
    \centering
    \includegraphics[scale=0.38]{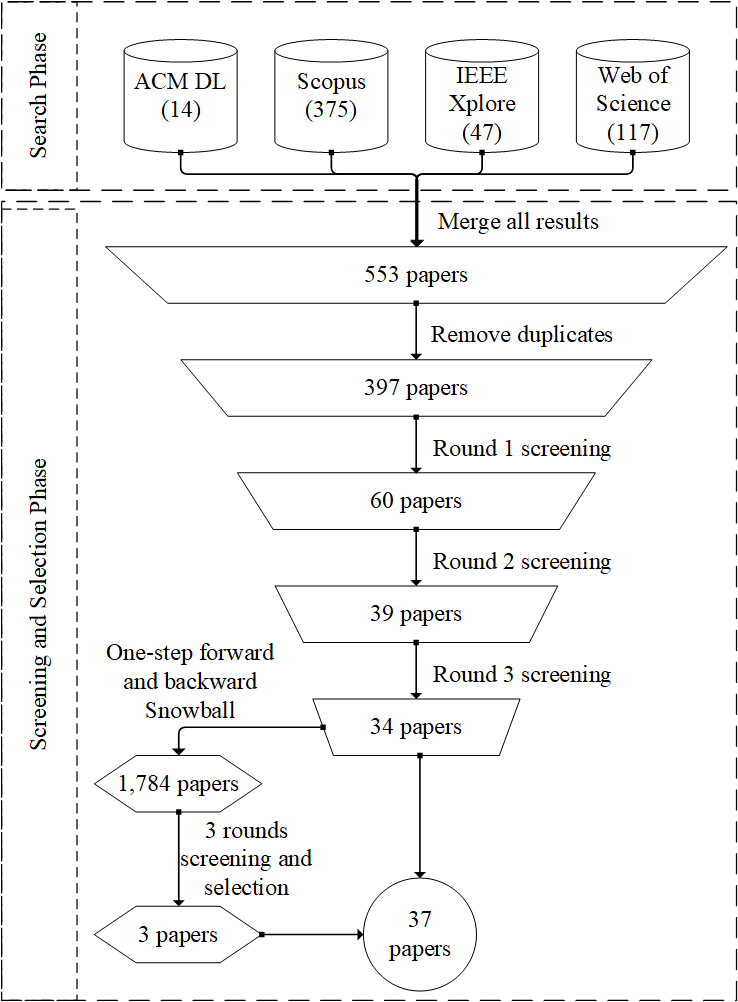}
    \caption{Details of study selection procedure following Kitchenham's guidelines \cite{kitchenham2004procedures}.}
    \label{fig:search}
\end{figure}

For the execution of this research, we utilized the systematic review management tool Covidence\footnote{\url{https://www.covidence.org/}}, which aids in organizing and streamlining the systematic approach applied during the selection phase of the studies. Figure \ref{fig:search} shows the details of the study selection procedure. By adhering to the search strategy detailed in the previous section, we explored the selected electronic databases and retrieved the papers. Initially, 553 papers were found, with some being duplicates due to overlapping sources in the databases. These papers underwent three rounds of the selection process described in the previous section, and 34 relevant papers were identified. Next, using this set of papers, we performed one-step backward and forward snowballing and identified 1,784 papers for screening. From this set, we first removed duplicates and applied our search query to filter irrelevant studies, resulting in 22 papers for further screening. Next, we followed three rounds of the selection process on this set, which resulted in 3 additional relevant papers. As an outcome, our final set of primary studies contained 37 papers, listed in Table \ref{tab:studies}. All searches were conducted in February 2025.

\subsubsection{Data extraction}
We used the data extraction form in Table~\ref{tab:extraction} to extract all the relevant information from the primary studies. Data fields F1-F6 are used to identify the paper and its bibliographic information. F5 refers to the type of paper as conference, journal, or workshop. For F6, we record the citation count for each paper according to the Semantic Scholar API. F7-F13 fields record the data for answering RQ1, and F14-F17 record the data for answering RQ2. F15 refers to the stages involved in analyzing app reviews, such as classifying or summarizing app reviews related to ethical concerns. For F17, we record the additional materials used in the analysis of app reviews, such as privacy policies or accessibility guidelines. F18 records the user-reported ethical concerns-related issues in the app reviews to answer RQ3. F19-F22 fields record the data to answer RQ4.

\begin{table}[]
    \centering
    \caption{Data extraction form}
    \label{tab:extraction}
    \begin{tabular}{lll}
        \textbf{Item ID} & \textbf{Field} & \textbf{Use} \\
        \hline
        F1 & Title & Documentation \\
        F2 & Author(s) & Documentation \\
        F3 & Year & Documentation \\
        F4 & Venue & Documentation \\
        F5 & Type & Documentation \\
        F6 & Citations & Documentation \\
        F7 & App Store(s) & RQ1 \\
        F8 & App Domain(s) & RQ1 \\
        F9 & Number of apps & RQ1 \\
        F10 & Number of app reviews & RQ1 \\
        F11 & Language of app reviews & RQ1 \\
        F12 & Terminology of app reviews & RQ1 \\
        F13 & App reviews collection method & RQ1 \\
        F14 & Study purpose & RQ2 \\
        F15 & Analysis stages & RQ2 \\
        F16 & Methodology & RQ2 \\
        F17 & Additional referenced materials & RQ2 \\
        F18 & Ethical concerns-related issues & RQ3 \\
        F19 & Results & RQ4 \\
        F20 & Limitations & RQ4 \\
        F21 & Future work & RQ4 \\
        F22 & Replication package & RQ4 \\
    \end{tabular}
\end{table}

\subsubsection{Information synthesis}
This phase entails the analysis and summarization of the 37 selected papers in our survey, which are presented and discussed in Section \ref{results}. Table \ref{tab:studies} showcases the final selection of papers included in this study. Here, we highlight the article identifier, title, authors, publication year, and the publication venue. The spreadsheet resulting from our data extraction and data grouping can be found in the supplementary material\footnote{\url{https://bit.ly/3GiTKli}}. We used this sheet to synthesize the extracted data for each RQ qualitatively. Since we aimed to summarize and interpret findings across studies rather than to develop a new theory, we did not apply a formal qualitative method such as thematic analysis. Instead, the analysis was carried out in a systematic but lightweight manner: data relevant to each RQ were organized according to the predefined data extraction fields and ethical concern categories, and reviewed iteratively to identify recurring patterns across the studies. The categorization was refined over multiple passes through the data, where emerging patterns were compared across studies and adjusted to improve consistency. Where applicable, findings were summarized in categorical or descriptive form to highlight recurring themes and distinctions across the literature. To improve reliability, the authors conducted repeated discussions to review intermediate categorizations, align interpretations with the extracted data, and resolve differences through consensus. While this process did not involve formal inter-coder agreement measures, it provided a structured mechanism to ensure consistency and reduce subjective bias in the synthesis.

\subsection{Reporting the review}
The data extracted from the primary studies were employed to address the four research questions. We closely followed Kitchenham's \cite{kitchenham2004procedures} guidelines when presenting the results.

    
    \begin{longtable}{lp{9cm}p{2cm}p{3cm}}
        \caption{List of primary studies selected in our survey.}
        \label{tab:studies}\\
        \textbf{ID} & \textbf{Title} & \textbf{Authors} & \textbf{Venue}\\
        \hline
        \hline
        S01 & ``for an App Supposed to Make Its Users Feel Better, It Sure is a Joke'' - An Analysis of User Reviews of Mobile Mental Health Applications & Haque et al. & HCI (2022)\\
        S02 & ``money Doesn't Buy You Happiness'': Negative Consequences of Using the Freemium Model for Mental Health Apps & Eagle et al. & HCI (2022)\\
        S03 & A Study of Gender Discussions in Mobile Apps & Shahin et al. & MSR (2023) \\
        S04 & Accessibility Feedback in Mobile Application Reviews: A Dataset of Reviews and Accessibility Guidelines & Reyes Arias et al. & CHI (2022)\\
        S05 & Analyzing Accessibility Reviews Associated with Visual Disabilities or Eye Conditions & Oliveira et al. & CHI (2023)\\
        S06 & Analyzing security issues of android mobile health and medical applications & Tangari et al. & JAMIA (2021)\\
        S07 & Analyzing user perspectives on mobile app privacy at scale & Nema et al. & ICSE (2022)\\
        S08 & Assessing Security, Privacy, User Interaction, and Accessibility Features in Popular E-Payment Applications & Kishnani et al. & EuroUSEC (2023)\\
        S09 & Automated detection, categorisation and developers' experience with the violations of honesty in mobile apps & Obie et al. & ESE (2023)\\
        S10 & AUTOREB: Automatically Understanding the Review-to-Behavior Fidelity in Android Applications & Kong et al. & SIGSAC-CCS (2015)\\
        S11 & Better addressing diverse accessibility issues in emerging apps: a case study using COVID-19 apps & Haggag et al. & MOBILESoft (2022)\\
        S12 & CHAMP: Characterizing Undesired App Behaviors from User Comments Based on Market Policies & Hu et al. & ICSE (2021)\\
        S13 & COVID-19 vs Social Media Apps: Does Privacy Really Matter? & Haggag et al. & ICSE-SEIS (2021)\\
        S14 & Discovering Privacy Harms from Education Technology by Analyzing User Reviews & Yang et al. & WPES (2024)\\
        S15 & Do Android App Users Care about Accessibility? An Analysis of User Reviews on the Google Play Store & Eler et al. & IHC (2019)\\
        S16 & Exploring the Influence of Software Evolution on Mobile App Accessibility: Insights from User Reviews & Oliveira et al. & JBCS (2024)\\
        S17 & Exploring User Perspectives of and Ethical Experiences With Teletherapy Apps: Qualitative Analysis of User  Reviews & Jo et al. & JMIR-MH (2023)\\
        S18 & Exploring user reviews to identify accessibility problems in applications for autistic users & Santiago et al. & JIS (2023)\\
        S19 & Fairness Concerns in App Reviews: A Study on AI-Based Mobile Apps	& Nasab et al. & TOSEM (2025)\\
        S20 & Finding the Needle in a Haystack: On the Automatic Identification of Accessibility User Reviews & AlOmar et al. & CHI (2021)\\
        S21 & Hark: A Deep Learning System for Navigating Privacy Feedback at Scale	& Harkous et al. & SP (2022)\\
        S22 & Identifying security issues for mobile applications based on user review summarization & Tao et al. & IST (2020)\\
        S23 & Identifying Users' Concerns in Lodging Sharing Economy Using Unsupervised Machine Learning Approach & Al-Ramahi et al. & ICDIS (2019)\\
        S24 & Learning Sentiment Analysis for Accessibility User Reviews & Aljedaani et al. & ASEW (2021)\\
        S25 & Privacy Analysis of Period Tracking Mobile Apps in the Post-Roe v. Wade Era & Dong et al. & ASE (2022)\\
        S26 & Privacy and Security Evaluation of Mobile Payment Applications Through User-Generated Reviews & Kishnani et al. & WPES (2022)\\
        S27 & Revealing the unrevealed: Mining smartphone users privacy perception on app markets & Hatamian et al. & COSE (2019)\\
        S28 & Short Text, Large Effect: Measuring the Impact of User Reviews on Android App Security \& Privacy & Nguyen et al. & SP (2019)\\
        S29 & The best ends by the best means: ethical concerns in app reviews & Tjikhoeri et al. & ESE (2024)\\
        S30 & The State of Accessibility in Blackboard: Survey and User Reviews Case Study & Aljedaani et al. & W4A (2023)\\
        S31 & Top concerns of user experiences in Metaverse games: A text-mining based approach & {\c{C}}all{\i} et al. & EC (2023)\\
        S32 & Uncovering Patterns in Users' Ethical Concerns About Software & Kara{\c{c}}am et al. & RE (2024)\\
        S33 & Understanding Security Issues based on App Comment Analysis & Chen et al. & DSA (2022)\\
        S34 & Unsupervised Summarization of Privacy Concerns in Mobile Application Reviews & Ebrahimi et al. & ASE (2022)\\
        S35 & Unveiling User Perspectives: Exploring Themes in Femtech Mobile App Reviews for Enhanced Usability and Privacy & Nellore et al. & HCI (2024)\\
        S36 & User Perspectives and Ethical Experiences of Apps for Depression: A Qualitative Analysis of User Reviews & Bowie-DaBreo et al. & CHI (2022)\\
        S37 & Using Machine Learning and Thematic Analysis Methods to Evaluate Mental Health Apps Based on User Reviews & Oyebode et al. & Acc. (2020)\\
        \hline
    \end{longtable}
    * For complete forms of venue names, see Table \ref{tab:venues}.

\begin{figure*}[htb]
    \centering
    \includegraphics[width=1\linewidth]{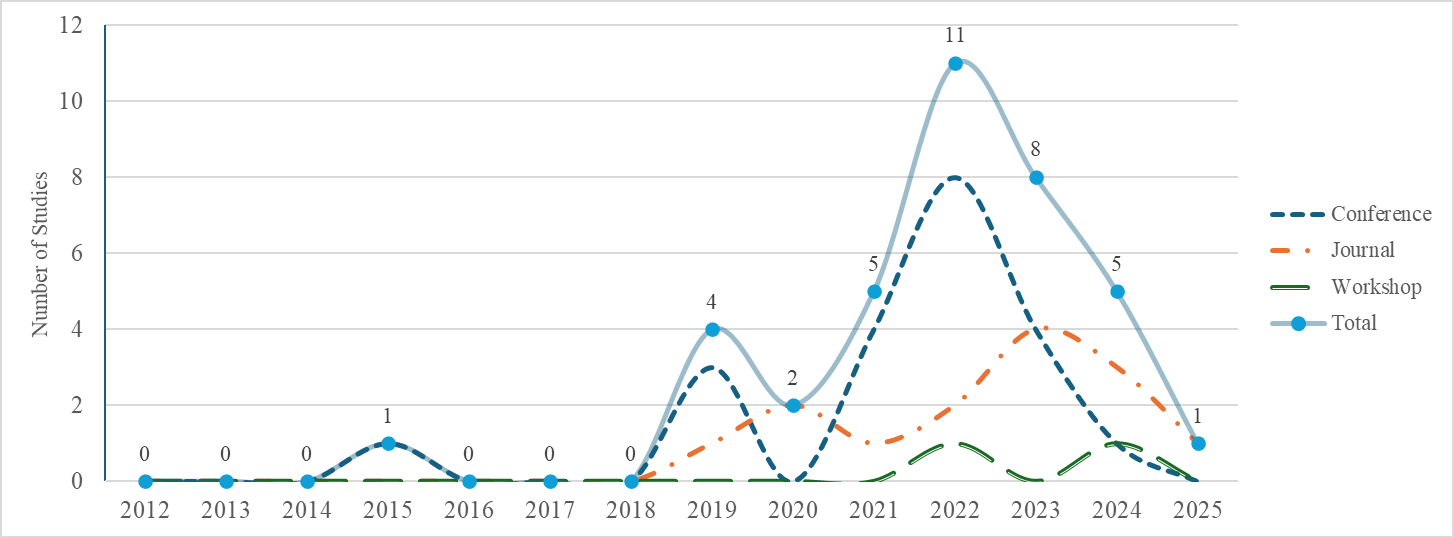}
    \caption{Temporal evolution of the number of studies in the domain of app review analysis for ethical concerns.}
    \label{fig:trends}
\end{figure*}

\begin{table*}[h]
    \centering
    \caption{Overview of publication venues.}
    \label{tab:venues}
    \begin{tabular}{p{13cm}p{2cm}}
        \textbf{Conferences, Journals, and Workshops} & \textbf{\# of primary studies} \\
        \hline
        \hline
        ACM Human-Computer Interaction (HCI) & 3\\
        ACM Transactions on Software Engineering and Methodology (TOSEM) & 1\\
        Brazilian Symposium on Human Factors in Computing Systems (IHC) & 1\\
        CHI Conference on Human Factors in Computing Systems (CHI) & 4\\
        Computers \& Security (COSE) & 1\\
        Conference on Computer and Communications Security (SIGSAC-CCS) & 1\\
        Empirical Software Engineering (ESE) & 2\\
        Entertainment Computing (EC) & 1 \\
        European Symposium on Usable Security (EuroUSEC) & 1\\
        IEEE Access (Acc.) & 1\\
        IEEE Symposium on Security and Privacy (SP) & 2\\
        Information and Software Technology (IST) & 1\\
        International Conference on Automated Software Engineering (ASE) & 2\\
        International Conference on Automated Software Engineering Workshops (ASEW) & 1\\
        International Conference on Data Intelligence and Security (ICDIS) & 1\\
        International Conference on Dependable Systems and Their Applications (DSA) & 1\\
        International Conference on Mining Software Repositories (MSR) & 1\\
        International Conference on Mobile Software Engineering and Systems (MOBILESoft) & 1\\
        International Conference on Software Engineering (ICSE) & 2\\
        International Conference on Software Engineering: Software Engineering in Society (ICSE-SEIS) & 1\\
        International Cross-Disciplinary Conference on Web Accessibility (W4A) & 1\\
        International Requirements Engineering Conference (RE) & 1\\
        Journal of Medical Internet Research Mental Health (JMIR-MH) & 1\\
        Journal of the American Medical Informatics Association (JAMIA) & 1\\
        Journal of the Brazilian Computer Society (JBCS) & 1\\
        Journal on Interactive Systems (JIS) & 1\\
        Workshop on Privacy in the Electronic Society (WPES) & 2\\
        \hline
    \end{tabular}
\end{table*}

\begin{figure*}[htb]
    \centering
    \includegraphics[width=1\linewidth]{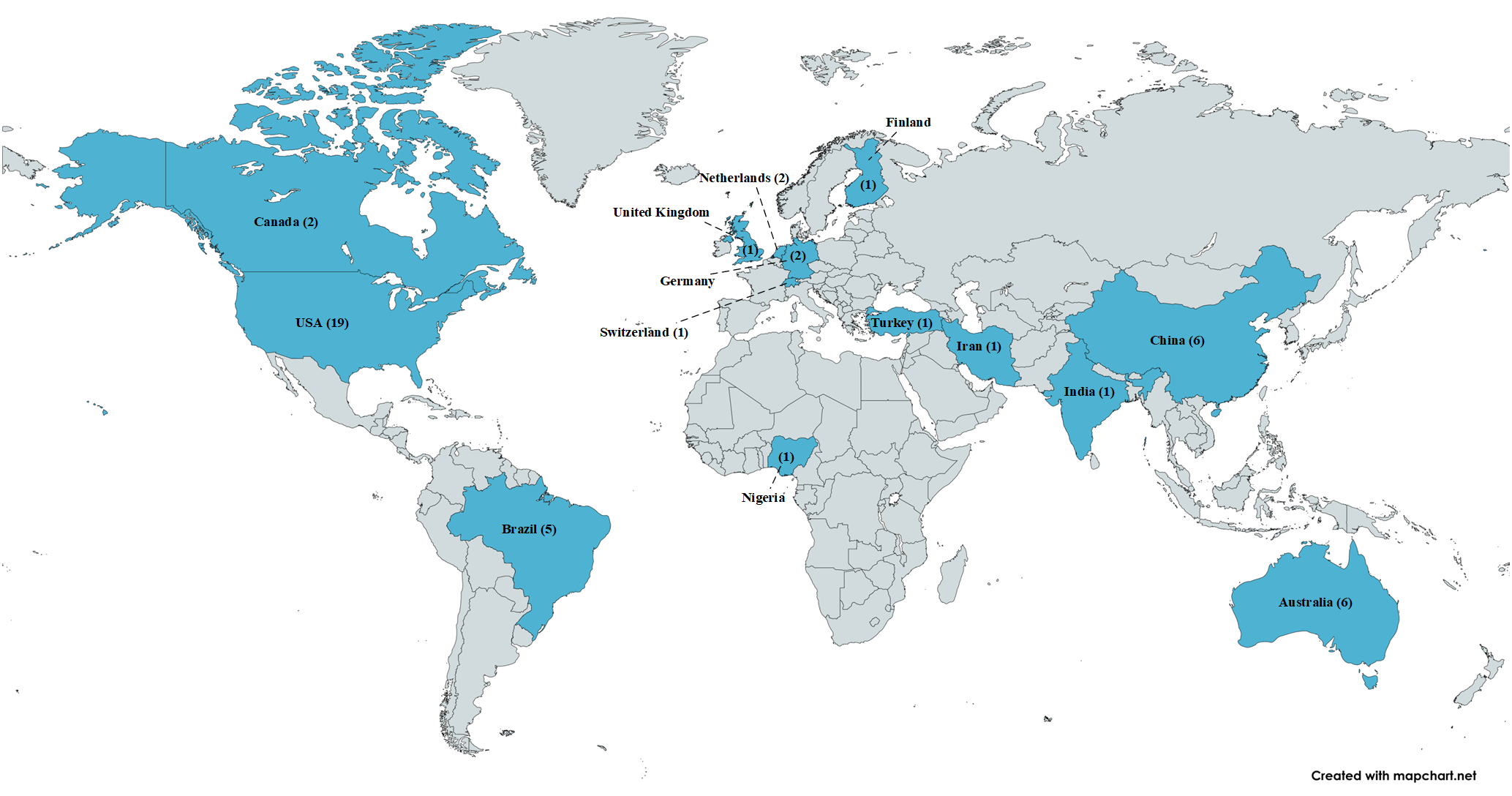}
    \caption{The geographical distribution of the papers published by teams on the subject.}
    \label{fig:map}
\end{figure*}

\begin{table}
    \centering
    \caption{Primary studies citation ranges.}
    \label{tab:cites}
    \begin{tabular}{p{3.5cm}p{3.5cm}p{6cm}}
        \textbf{\#Citations range} & \textbf{\#primary studies} & \textbf{Study IDs} \\
        \hline
        \hline
        0-10 & 16 & S16, S35, S23, S33, S14, S32, S19, S18, S17, S29, S03, S09, S08, S26, S02, S34 \\
        11-20 & 8 & S30, S11, S04, S05, S12, S25, S31, S24 \\
        21-30 & 5 & S01, S06, S13, S36, S21 \\
        31-50 & 6 & S07, S22, S15, S20, S27, S28 \\
        >50 & 2 & S10, S37 \\
        \hline
    \end{tabular}
\end{table}

\begin{table}
    \centering
    \caption{Authors' affiliations.}
    \label{tab:aff}
    \begin{tabular}{p{4cm}p{9cm}}
        \textbf{Authors' Affiliation} & \textbf{Studies} \\
        \hline
        \hline
        Academic & S01, S02, S03, S04, S05, S06, S08, S11, S12, S13, S14, S15, S16, S17, S18, S19, S20, S22, S23, S24, S25, S26, S27, S28, S29, S30, S31, S32, S33, S34, S35, S37 \\
        Collaboration & S09, S10, S36\\
        Industrial & S07, S21\\
        \hline
    \end{tabular}
\end{table}

\section{Result Analysis} \label{results}
In this section, we first provide a broad summary of the trends in publications, then analyze the synthesized data from the primary studies to answer our research questions.

\subsection{\textbf{Publication Trends}}
This section presents the trends in the primary studies based on publication year, publication venue, venue type, authors' collaboration, and citations.

In recent decades, there has been a noticeable but modest increase in the number of publications focusing on app review analysis for exploring users' ethical concerns. Figure \ref{fig:trends} presents the distribution of primary studies from 2012 to 2025, categorized by venue type: Conference, Journal, and Workshop, with a total count also provided. The Y-axis indicates the number of studies, while the X-axis marks the year of publication. It is evident that prior to 2019, there was a singular publication (S10 - a conference paper) in 2015 that addressed the analysis of privacy and security issues derived from app reviews. This particular study has established itself as the seminal work for privacy and security analysis, as it is referenced by the majority of the primary studies within this field.

The number of studies increased in 2019 (4 total), consisting of three conference papers (S15, S23, S28) and one journal paper (S27). The highest number of publications was observed in 2022, with 11 total studies (S01, S02, S04, S07, S11, S21, S25, S26, S33, S34, S36), out of which one was a workshop paper (S26) and two were journal papers (S01, S02). The surge in studies from 2019–2022 reflects a growing interest in this domain, possibly following the development of new technologies or pressing research problems. The prominence of conference publications during these peak years highlights a swift dissemination of initial findings and highlights the significance of collaborative research sharing within the community. From 2022 onwards, a declining trend is observed: 8 studies (S03, S05, S08, S09, S17, S18, S30, S31) in 2023, and 5 (S14, S16, S29, S32, S35) in 2024. Workshop contributions remained minimal throughout this timeframe, with one publication recorded in each of the years 2022 (S26) and 2024 (S14).

Certain publication venues demonstrate a significant frequency in disseminating research within this domain. We present an overview of the publication venues associated with the primary studies in Table~\ref{tab:venues}. Our analysis highlights that the 37 primary studies are allocated across 28 distinct publication venues, with the majority of these venues having published one to two papers on the topic. The venues specializing in Human-Computer Interaction research seem to be the favored publication outlets in this area, contributing three papers (S01, S02, S35) to ``ACM Human-Computer Interaction (HCI)'' and four papers (S04, S05, S20, S36) to the ``CHI Conference on Human Factors in Computing Systems (CHI)''.

Table~\ref{tab:aff} shows the affiliation of the authors with their respective studies, where 32 studies are academic-affiliated, three studies are collaboration-affiliated (S09, S10, S36), and only two studies (S07, S21) are industry-affiliated.

The map in Figure \ref{fig:map} depicts the graphical distribution of teams, with the highest number of studies published by the authors from the United States team (19 studies). Note that we included the country of all authors in the analysis. Among all studies, 8 studies (S03, S05, S07, S09, S12, S16, S19, S30) had authors from more than one country.

We also analyzed the impact of studies in terms of their citation count as shown in Table \ref{tab:cites}. The average number of citations is 20, with studies S37 and S10 receiving the highest, with a count of 108 and 69 citations, respectively. Note that the citation count is fetched using the SemanticScholar API, which may change over time.

\subsection{RQ1: App Reviews Characteristics}
This subsection elaborates on RQ1: What are the attributes of the collected app reviews, and what techniques were used in their collection?\\

\noindent\textbf{Dataset size. }
The datasets utilized in primary studies exhibited substantial variability in terms of the number of apps and app reviews evaluated. The range of mobile apps spanned from a singular app (e.g., Blackboard in S30 and Airbnb in S23) to an excess of 1.3 million apps (S21). The majority of studies concentrated on smaller dataset sizes, generally falling between 10 and 713 apps, with a limited number of exceptions scrutinizing thousands or millions of apps (S06, S07, S10, S12, S21, S28). The number of app reviews examined varied considerably, from 500 (S23) to an astounding 626 million (S21) reviews. Furthermore, the initial number of app reviews collected and the subset of relevant reviews analyzed, following specific filtering criteria, demonstrated significant variability. For instance, S03 collected 7 million app reviews and subsequently analyzed only 1,440 relevant reviews; S07 compiled 287 million reviews and identified approximately 440,000 relevant reviews from this dataset.

We also observed that the number of apps does not always correlate with the number of app reviews. For example, S25 collected app reviews from 35 apps and analyzed only 268 relevant reviews, whereas S30 collected app reviews from just one app and analyzed 7,534 reviews. This substantial disparity in the number of app reviews compiled and analyzed, as well as in the number of apps and app reviews, was observed due to the specific focus area of the primary studies; for example, S25 had a specific focus on privacy analysis of period tracking apps in the post-Roe vs Wade era.\\

\noindent\textbf{App stores. }
The Google Play Store emerged as the most frequently utilized app marketplace in the primary studies for the purpose of collecting app reviews. A number of studies (S01, S09, S11, S13, S14, S17, S34, S35, S36, S37) additionally employed the Apple App Store to perform cross-platform analyses, whereas S23 solely focused on the Apple App Store. Furthermore, S36 specifically examined the UK versions of both stores in their research. S20 included the Amazon App Store in their analysis alongside the Google Play Store. S12, in addition to the Google Play Store, took into account eight Chinese app markets, namely 360 Market, Huawei, Lenovo, Meizu, OPPO, Vivo, Xiaomi, and Tencent Myapp. S04 utilized FDroid\footnote{\url{https://f-droid.org/}}, a repository for Free and Open Source Software (FOSS) apps on Android, to gather open source apps and subsequently collected the relevant app reviews from the Google Play Store.\\

\noindent\textbf{App domains. }
Several primary studies have analyzed app reviews within a particular app domain, with the mental health domain being the primary focus of studies S01, S02, S17, S34, S36, and S37. Studies S08 and S26 concentrated on the finance domain, while S11 and S13 investigated COVID-19 apps. The education sector was explored by three studies, namely S14, S18, and S30. Study S23 analyzed app reviews of lodging sharing economy apps, and S25 exclusively assessed period tracking apps. Study S31 specifically focused on metaverse apps, whereas S34 investigated apps in the investing and food delivery domains.\\

\noindent\textbf{App reviews language. }
Most of the studies analyzed only English language reviews, except only four studies (S11, S12, S13, and S18) that considered additional languages. S12 analyzed Chinese reviews in addition to English reviews, whereas S18 only considered Portuguese reviews. S11 and S13 considered app reviews of different languages from 30 countries.\\

\noindent\textbf{App reviews collection method. }
The app review collection techniques used in the surveyed studies can be broadly classified into five major categories: custom web scraping, use of public or unofficial APIs, manual extraction, reuse of existing datasets, and automated platforms or tools.
\begin{itemize}
    \item \textit{Custom web scraping tools}: Several studies (e.g., S01, S10, S14, S17, S22, S28, S34) built Python crawlers tailored to specific app stores, often using multiple IP addresses to bypass anti-scraping barriers. These scrapers typically targeted apps with minimum thresholds (e.g., \(\geq\)100 downloads or \(\geq\)50 characters in reviews), and often filtered for English-language reviews to ensure consistency in text analysis.
    \item \textit{Public or unofficial APIs}: Many studies leveraged open-source or unofficial APIs for efficient and scalable data collection. Specifically, S05, S08, S13, S15, S26, S27, and S31 utilized tools such as the google-play-scraper\footnote{\url{https://github.com/JoMingyu/google-play-scraper}}\footnote{\url{https://github.com/facundoolano/google-play-scraper}} and the google-play-api\footnote{\url{https://github.com/facundoolano/google-play-api}}, allowing for the aggregation of thousands to millions of reviews per app. These libraries frequently included integrated functionalities, such as language filtering and relevance sorting, and were extensively utilized for comprehensive data analysis.
    \item \textit{Manual review collection}: A smaller set of studies (S02, S18, S23, S36) manually curated reviews, typically to ensure representative samples (e.g., equal positive and negative reviews or most helpful and most critical reviews) or to focus on specific app categories (e.g., educational apps in Portuguese) where automated tools were less suitable.
    \item \textit{Reuse of existing datasets}: Several studies relied on previously published datasets, especially when examining ethical concerns such as accessibility and fairness. For example, S04, S16, S20, S24, S32, and S33 reused datasets from foundational works S15, S09 or S05. These datasets were typically filtered by keyword or manually validated for relevance, allowing new research to build upon validated and pre-processed app review datasets.
    \item \textit{Automated platforms and monitoring tools}: A few studies used third-party platforms and monitoring services to automate review collection and metadata scraping. Notable among them are S12, which used the Kuchuan\footnote{\url{https://www.kuchuan.com/}} platform for daily data pulls, and S35, which employed SerpApi\footnote{\url{https://serpapi.com/}} for structured metadata extraction alongside reviews. S37 used Heedzy\footnote{\url{https://heedzy.com/}}, a review scraping tool that facilitated large-scale review downloads with minimal manual effort.
\end{itemize}

Some studies used a combination of the above methods, such as S11 mixed automated scraping with manual validation. Additionally, a few studies (S07, S21, S25, and S30) did not specify their collection methods but likely used internal tools or APIs, given their scale and review count.\\

\noindent\textbf{Terminology of app reviews. }
\begin{figure}[htb]
    \centering
    \includegraphics[scale=0.65]{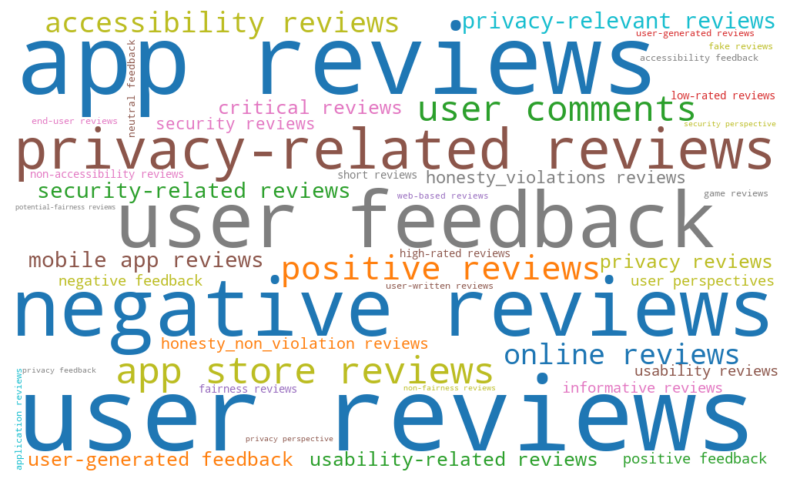}
    \caption{Word cloud of the terminologies used in the primary studies to refer to ``app reviews''.}
    \label{fig:terms}
\end{figure}
Figure \ref{fig:terms} illustrates the word cloud representing the terminologies utilized in the primary studies to denote ``app reviews.'' The terms ``user reviews,'' ``app reviews,'' and ``user feedback'' appear most frequently, indicating that these are the conventional or default expressions employed across the majority of studies. Their prevalence signifies a shared understanding in referring to app-based textual evaluations from users. Specialized phrases such as ``privacy reviews,'' ``accessibility reviews,'' and ``security reviews'' suggest that researchers often customize their terminology according to the specific ethical concern being examined. Terminologies such as ``positive reviews,'' ``negative reviews,'' ``critical reviews,'' ``relevant reviews,'' and ``informative reviews'' are commonly used to signify sentiment polarity or the perceived value of the review content. This diversity in terminology reflects both domain-specific objectives (e.g., accessibility or privacy) and characteristics of the data (e.g., rating level, review length, review source).\\

\begin{tcolorbox}[breakable,colback=gray!10!white,colframe=gray!75!black,title=Summary of RQ1]
The dataset used by primary studies varied significantly in terms of the number of apps (single app to 1.3 million apps) and the number of app reviews (500 reviews to 287 million reviews). The Google Play Store was most utilized one to collect app reviews, with the Apple App Store being the second most used. Most of the studies did not focus on any specific app domain, while among domain-specific studies, the mental health domain was the primary focus.

Most of the studies considered only English language reviews, with only four studies analyzing non-English reviews. Primary studies used different techniques to collect app reviews, which can be categorized into five major categories, including custom web scraping, use of public or unofficial APIs, manual extraction, reuse of existing datasets, and automated platforms or tools. Primary studies most commonly used ``user reviews'', ``app reviews'', and ``user feedback'' terms to refer to the app reviews.
\end{tcolorbox}

\subsection{RQ2: Research Methods}
This subsection elaborates on RQ2: What objectives, methods, and resources are frequently utilized to analyze app reviews related to ethical concerns?\\

\begin{table}[t]
    \caption{Categorization of primary studies into the ethical concerns based on their purpose of analysis (detailed in \ref{apndxAA}) and the user-reported issues discussed from the app reviews (discussed in Section \ref{issues}).}
    \label{tab:concerns}
        \centering
        \begin{center}
    \begin{tabular}{m{1.6cm}|m{1cm}m{1.1cm}m{1.9cm}m{2.1cm}m{1.1cm}m{2.3cm}m{1cm}}
         \textbf{Study ID} & \textbf{Privacy} & \textbf{Security} & \textbf{Accessibility} & \textbf{Transparency} & \textbf{Fairness} & \textbf{Accountability} & \textbf{Safety} \\
         \hline
         S01, S23 & \faIcon[]{check} &   &   &  \faIcon[]{check} &  \faIcon[]{check} &  \faIcon[]{check} &  \faIcon[]{check} \\
         \hline
         S02, S09 &   &   &   &  \faIcon[]{check} &  \faIcon[]{check} &  \faIcon[]{check} &   \\
         \hline
         S03 &   &   &   &  \faIcon[]{check} &  \faIcon[]{check} &  \faIcon[]{check} &  \faIcon[]{check} \\
         \hline
         S04, S05, S15, S16, S18, S20, S24, S30 &   &   &  \faIcon[]{check} &   &   &   &   \\
         \hline
         S06, S10, S22, S25, S27, S28, S33, S34 &  \faIcon[]{check} &  \faIcon[]{check} &   &   &   &   &   \\
         \hline
         S07, S21 &  \faIcon[]{check} &   &   &   &   &   &   \\
         \hline
         S08, S26 &  \faIcon[]{check} &  \faIcon[]{check} &  \faIcon[]{check} &   &   &   &   \\
         \hline
         S11, S13 &  \faIcon[]{check} &   &  \faIcon[]{check} &   &   &   &   \\
         \hline
         S12 &  \faIcon[]{check} &  \faIcon[]{check} &   &  \faIcon[]{check} &   &   &  \faIcon[]{check} \\
         \hline
         S14 &  \faIcon[]{check} &  \faIcon[]{check} &   &   &   &   &  \faIcon[]{check} \\
         \hline
         S17, S19 &   &   &   &  \faIcon[]{check} &  \faIcon[]{check} &   &   \\
         \hline
         S29, S32, S36, S37 &  \faIcon[]{check} &  \faIcon[]{check} &  \faIcon[]{check} &  \faIcon[]{check} &  \faIcon[]{check} &  \faIcon[]{check} &  \faIcon[]{check} \\
         \hline
         S31 &  \faIcon[]{check} &  \faIcon[]{check} &   &   &  \faIcon[]{check} &  \faIcon[]{check} &  \faIcon[]{check} \\
         \hline
         S35 &  \faIcon[]{check} &  \faIcon[]{check} &  \faIcon[]{check} &   &   &  \faIcon[]{check} &   \\
         \hline
    \end{tabular}
    \end{center}
\end{table}

\noindent\textbf{Study objectives. }
The primary studies represented a wide range of analytical objectives centered on analyzing ethical concerns from app reviews. We categorized these studies based on their purpose into a previously identified set of common ethical concerns as shown in Table \ref{tab:concerns}.

Many studies (S01, S06, S07, S08, S10, S11, S12, S13, S14, S21, S22, S23, S25, S26, S27, S28, S29, S32, S33, S34, S35, S36, S37) focused on analyzing users' privacy and security concerns related to data privacy, security vulnerabilities, and surveillance risks. These analyses often complemented automated assessments and were used to propose improvements for privacy-preserving app design or highlight regulatory challenges. Additionally, most studies analyzed privacy and security concerns together due to their interwoven nature \cite{wright2011framework}, except a few studies (S07 and S21) that explicitly mentioned excluding security concerns in their analysis.

Studies S04, S05, S08, S11, S13, S15, S16, S18, S20, S24, S29, S30, S32, S35, S36, and S37 explored accessibility concerns in depth, examining how people with disabilities or specific needs (e.g., autism) experience mobile apps. These studies often aligned app reviews with established accessibility guidelines, such as Web Content Accessibility Guidelines (WCAG) or Guidelines for Accessible Interfaces for Autistics (GAIA), and aimed to inform inclusive design practices.

Studies S01, S02, S03, S09, S12, S17, S19, S23, S29, S31, S32, S36, and S37 uncovered transparency and fairness-related concerns to encourage the development of gender-inclusive mobile apps and fair and transparent subscription models in mobile apps. We also observed that transparency and fairness concerns often occurred together because any non-transparent functionality of an app, such as a non-transparent subscription model, was also perceived as unfair by the users.

A few domain-specific studies (S02, S23, S29, S31, S32, S36, and S37) analyzed accountability and safety concerns with the purpose of highlighting issues related to external factors of apps, such as the lack of customer support, unsafe riders, and harmful communities.

We provide a detailed explanation of the purposes extracted from the primary studies in \ref{apndxAA}.\\

\noindent\textbf{Analysis stages and applied methods. }
All 37 primary studies conducted empirical investigations utilizing app reviews, with most delivering both quantitative and qualitative findings, while a few focused primarily on quantitative outcomes (S10, S12, S22, S24). The examination followed a multistage methodology encompassing seven fundamental stages.

\begin{itemize}
    \item \textit{Data collection and filtering}: Reviews were extracted from Google Play and the Apple App Store using explicit inclusion criteria. Filtering was applied based on length (S01, S02, S14, S19, S26, S30, S34--S36), date (S01, S02, S25), language (S19, S22, S26, S30, S35), star ratings (S14, S31), and sentiment (S01, S02, S08, S22, S26, S27, S33). Keyword filtering drew on predefined sets from authoritative references (S03--S09, S15, S17--S19, S22, S26--S29, S34, S35) such as WHO documents\footnote{\url{https://icd.who.int/browse10/2016/en\#/VII}} or BBC Mobile Accessibility Guidelines\footnote{\url{https://www.bbc.co.uk/accessibility/forproducts/guides/mobile/}}, supplemented by keyword co-occurrence analysis (S10) or KeyBERT \cite{maarten2021} (S03, S19). One study (S21) applied natural language inference for filtering.

    \item \textit{Data preprocessing and cleaning}: Standard techniques included tokenization (S20, S24, S27, S28), lemmatization (S20, S24, S27, S28, S31, S32, S34, S37), stop word removal (S14, S20, S23, S24, S27, S29, S31, S37), punctuation and emoji elimination (S14, S20, S22, S24, S26, S29--S32, S34, S36, S37), case normalization (S24, S29, S31, S37), and contraction expansion (S22, S33, S37). Additional techniques included part-of-speech tagging (S32), slang replacement (S37), word segmentation (S22, S33), and spell correction (S22, S26, S33) using libraries such as PyEnchant\footnote{\url{https://pyenchant.github.io/pyenchant/}} or language-tool-python\footnote{\url{https://github.com/jxmorris12/language_tool_python}}.

    \item \textit{Feature extraction and representation}: Textual data was converted into machine-learning-ready formats using TF-IDF \cite{ramos2003using} (S10, S12, S19, S23, S24, S27, S29, S35, S37), Word2Vec \cite{church2017word2vec} (S03, S19), Universal Sentence Encoder \cite{cer2018universal} (S07, S19), and GloVe \cite{pennington2014glove} (S27, S34). Some studies applied feature hashing \cite{weinberger2009feature} (S20) or mutual information \cite{torkkola2003feature} (S34) to identify predictive terms.

    \item \textit{Classification and labeling}: Reviews were categorized via manual coding and inductive thematic analysis \cite{fereday2006demonstrating} (S01, S02, S04--S06, S08, S16--S18, S22--S26, S31, S34--S36), supervised machine learning (S03, S07, S09, S10, S19--S21, S27--S29, S33, S37), or semi-automated keyword-driven classification (S11--S15, S26, S30). Algorithms included Logistic Regression, SVMs, Random Forest, XGBoost, CNNs, LSTMs, and RoBERTa \cite{liu2019roberta}. Inter-rater reliability was reported using Cohen's \(\kappa\) \cite{fleiss1973equivalence}, and multi-label classifiers were applied where reviews addressed multiple concerns (S10, S29).

    \item \textit{Validation and triangulation}: In thematic studies, reliability was established through coder agreement and consensus meetings (S03, S05, S12, S16--S19, S30). Machine learning studies reported precision, recall, F1-score, and AUC via cross-validation (S03, S07, S09, S10, S12, S19--S21, S24, S26--S29, S33, S34, S37). Some studies triangulated qualitative themes with app usage audits (S02), developer interviews (S09), and traffic analysis (S25, S27), with manual inspection further supporting automated findings (S13, S25, S31).

    \item \textit{Interpretation and synthesis}: Findings were mapped to established guidelines such as WCAG or biomedical ethical principles (S05, S16, S17, S30, S36), supported by sentiment and trend analyses (S05, S21, S24, S25, S35, S36). Actionable implications addressed user requirements (S16, S23, S26), developer gaps (S09, S11, S28), and policy gaps (S13, S25). Some studies additionally developed tools or models (S11) or conducted cross-sectional comparisons across time (S28, S35) or app versions (S28).
\end{itemize}

\noindent\textbf{Additional referenced materials. }
In addition to app reviews, the studies also utilized various tools and frameworks that helped with various research activities. These materials are outlined below:
\begin{itemize}
    \item \textit{Ethical guidelines and frameworks}: Primary studies utilized ethical guidelines and frameworks to identify the violation of the guidelines by mapping the relevant app reviews to them. These guidelines were also used to develop a comprehensive set of keywords to extract the relevant reviews. We identified the following ethical guidelines and frameworks from the primary studies: Web Content Accessibility Guidelines (WCAG) 2.1\footnote{\url{https://www.w3.org/TR/WCAG21/}} / 2.2\footnote{\url{https://www.w3.org/TR/WCAG22/}} (S04, S16), BBC Mobile Accessibility Guidelines\footnote{\url{https://www.bbc.co.uk/accessibility/forproducts/guides/mobile/}} (S04, S05, S15-S17, S30), Google Material Design\footnote{\url{https://m3.material.io/}} (S05, S16), Guidelines for Accessible Interfaces for Autistics (GAIA)\footnote{\url{https://github.com/talitapagani/gaia}} (S18), Solove’s taxonomy of privacy violations \cite{solove2005taxonomy} (S21), Wang's taxonomy of privacy enhancing technologies \cite{wang2009privacy} (S21), General Data Protection Regulation (GDPR) guidelines\footnote{\url{https://gdpr-info.eu/}} (S08), California Consumer Privacy Act (CCPA)\footnote{\url{https://www.oag.ca.gov/privacy/ccpa}} (S08), American Psychological Association (APA) Ethical Principles\footnote{\url{https://www.apa.org/ethics/code}} (S17), and App Privacy Policies (S08, S12, S13, S25, S35) and Terms of Service (S02, S13).
    \item \textit{Static and dynamic app analysis tools}: We identified the following static and dynamic app analysis tools that were used by the primary studies to perform static code analysis, dynamic analysis of APK files, vulnerability scanning, API usage analysis, embedded libraries analysis, and network analysis:  MobSF \footnote{\url{https://github.com/MobSF/Mobile-Security-Framework-MobSF}} (S08), AndroBugs \footnote{\url{https://github.com/AndroBugs/AndroBugs_Framework}} (S08), RiskInDroid \footnote{\url{https://github.com/ClaudiuGeorgiu/RiskInDroid}} (S08), LibRadar \footnote{\url{https://github.com/pkumza/LibRadar}} (S12), FlowDroid \footnote{\url{https://github.com/secure-software-engineering/FlowDroid}} (S12), AppOpsManager\footnote{\url{https://developer.android.com/reference/android/app/AppOpsManager}} (S27), mitmproxy\footnote{\url{https://mitmproxy.org/}} (S25), LibScout\footnote{\url{https://github.com/reddr/LibScout}} (S28), PScout\footnote{\url{https://security.csl.toronto.edu/pscout/}} (S25), and Testin\footnote{\url{https://testin.net/}} (S12). The analysis using these tools was utilized to validate users' concerns and identify the gap between actual app behaviors and users' perceptions.
    \item \textit{Supplementary data sources}: A few primary studies also included perspectives of other stakeholders such as app developers and app owners, and app metadata and demo screenshots to validate users' concerns: developer interviews (S09, S33), app store metadata (S06), app descriptions or screenshots (S01, S11), user interviews and surveys (S30), and developer replies on app stores (S28).
\end{itemize}

\begin{tcolorbox}[breakable,colback=gray!10!white,colframe=gray!75!black,title=Summary of RQ2]
The objectives of the primary studies varied significantly, ranging from the manual or thematic analysis of app reviews to the automated classification and clustering of reviews. However, these objectives were centered around addressing users' concerns related to privacy, security, accessibility, transparency, fairness, accountability, and safety. Furthermore, these studies employed multiple stages of analysis and used different sets of techniques to analyze ethical concern-related app reviews. These stages can be categorized into seven fundamental stages: data collection and filtering, data preprocessing and cleaning, feature extraction and representation, classification and labeling, validation and triangulation, and interpretation and synthesis. The primary studies also utilized additional material in their analysis, including ethical guidelines and frameworks, static and dynamic app analysis tools, and supplementary data sources (e.g., app store metadata).
\end{tcolorbox}

\subsection{RQ3: User-reported Ethical Concerns-related Issues} \label{issues}
This subsection elaborates on RQ3: What particular ethical concerns-related issues do users report in the analyzed app reviews? From primary studies, we identified significant mentions of ethical concerns-related issues reported by users in app reviews and categorized them into a previously defined set of ethical concerns.\\

\noindent\textbf{Privacy-related issues. }Privacy violations represent a significant and widespread concern, as users consistently report unauthorized data access, data misuse, and fear of surveillance. Many mobile apps collect sensitive data without user consent, retain such data for extended periods beyond what is necessary, or share it with third parties, often disguised under ambiguous policy statements. In the following, we outline a few privacy-related issues extracted by the primary studies from app reviews and include the complete list of issues in the \ref{apndxA}.
\begin{itemize}
    \item \textit{The user was terrified since their encounter with the app seemed like they were being monitored.} (S01)
    \item \textit{Location permission is the most discussed permission, appearing in over 40,000 reviews.} (S07)
    \item \textit{User was not provided with an alternative layer of authentication, and got locked out of their account.} (S08)
    \item \textit{Collect user’s current address and places they regularly visit, businesses they use, and people they are close by.} (S13)
    \item \textit{Collection of data, such as browsing history, device information, authentication tokens (such as photo ID shown before taking exams), and financial information (e.g., credit card number used to pay for the service remotely).} (S14)
\end{itemize}

\noindent\textbf{Security-related issues. }Security concerns focus on malware, data leaks, phishing, and unauthorized access. Users highlight vulnerabilities such as a lack of encryption, account hacks, and security bugs, leading to financial fraud and loss of sensitive information. In the following, we outline a few security-related issues extracted by the primary studies from app reviews and include the complete list of issues in the \ref{apndxA}.
\begin{itemize}
    \item \textit{Explicit mentions of malware or suspicious activity.} (S06)
    \item \textit{Illegal background behavior (e.g., SMS).} (S12)
    \item \textit{Intrusive access to the device and peripherals.} (S14)
    \item \textit{Invasive, parasitic malware.} (S22)
    \item \textit{Difficult or impossible to revert wrong transactions.} (S26)
    \item \textit{Malwares and computer viruses.} (S31)
\end{itemize}

\noindent\textbf{Accessibility-related issues. }Accessibility failures leave users, especially those with disabilities, unable to use apps effectively. Issues include poor UI design, lack of compatibility with assistive tech, and exclusionary defaults (e.g., no captions, screen reader issues, small fonts). In the following, we outline a few accessibility-related issues extracted by the primary studies from app reviews and include the complete list of issues in the \ref{apndxA}.
\begin{itemize}
    \item \textit{Reviews on the necessity of increasing font size, as well as updating the apps for zoom features.} (S05)
    \item \textit{Difficult for users to navigate, locate desired features, and complete transactions efficiently.} (S08)
    \item \textit{Apps are only accessible by users using a local country app store and they are not available in the international store.} (S11)
    \item \textit{Using Bluetooth for tracing, which affects the connectivity with other devices such as headphones and smartwatches.} (S13)
    \item \textit{Users mentioned issues with theme/mode, zoom, customization, media, font, contrast, impairment, flickering, and language.} (S15)
\end{itemize}

\noindent\textbf{Transparency-related issues. }Lack of transparency in data practices, fees, and app capabilities undermines user trust. Users often face hidden costs, misleading claims, or discover after installation that the app behaves differently from its description. In the following, we outline a few transparency-related issues extracted by the primary studies from app reviews and include the complete list of issues in the \ref{apndxA}.
\begin{itemize}
    \item \textit{It often took a lot of time and effort on the user end to understand that a particular functionality in the app that attracted her is not available in the free version.} (S01)
    \item \textit{Users also felt that payment plan descriptions were imprecise, leaving them unclear about which features were offered in the free version, and confused about the charging model.} (S02)
    \item \textit{Users found it inaccurate or less useful when the translation feature does not provide any clue of gender.} (S03)
    \item \textit{Users perceive the cancellation and refund policy as unfair, nontransparent, or deliberately misleading.} (S09)
    \item \textit{Frustrated as the cost was not disclosed until they finished the sign-up process, wishing pricing information was clear up front.} (S17)
\end{itemize}

\noindent\textbf{Fairness-related issues. }Users cite discrimination, algorithmic bias, and economic inequity. Apps often favor one demographic over others, apply unfair rules, or offer unequal experiences across platforms or languages. In the following, we outline a few fairness-related issues extracted by the primary studies from app reviews and include the complete list of issues in the \ref{apndxA}.
\begin{itemize}
    \item \textit{Users have occasionally expressed their disappointments with moderators ``feeling entitled'' and ``abusing their power'' by showing conscious biased behavior.} (S01)
    \item \textit{Several users raised their concerns as apps did not offer the option to choose between male and female voices when interacting with bots.} (S03)
    \item \textit{Users complain of unfairness in either the process or outcome of the app.} (S09)
    \item \textit{The subscription model adopted by some apps resulted in them paying fees that were not commensurate with the services that they received.} (S17)
    \item \textit{Receiving different quality of features and services in different platforms and devices.} (S19)
\end{itemize}

\noindent\textbf{Accountability-related issues. }Accountability issues arise from unresponsive customer support, lack of enforcement of ethical standards, and absence of proper redressal for harms caused by the app or its community. In the following, we outline a few accountability-related issues extracted by the primary studies from app reviews and include the complete list of issues in the \ref{apndxA}.
\begin{itemize}
    \item \textit{The apps frequently failed to provide support during emergency in a timely manner.} (S01)
    \item \textit{Users expected better moderation from the app owners to remove these contents regularly, ban the accounts, or facilitate reporting such behaviours.} (S03)
    \item \textit{Users have informed the issues experienced while engaging customer support teams for complaint resolution.} (S08)
    \item \textit{Users perceive that the app provides false or inaccurate information.} (S09)
    \item \textit{No policies in place to protect their guests against a case manager who isn’t properly investigating their case.} (S23)
\end{itemize}

\noindent\textbf{Safety-related issues. }Safety risks include harm from app content (e.g., pro-self-harm communities), predatory users, and physical risks (e.g., unsafe rides, stalker access). In the following, we outline a few safety-related issues extracted by the primary studies from app reviews and include the complete list of issues in the \ref{apndxA}.
\begin{itemize}
    \item \textit{Trolling, unpleasant behavior, and harassment were all common, along with negative comments about race, gender, and religion.} (S01)
    \item \textit{Communities on apps (e.g., social media, gaming groups) are violent or share hateful comments against specific genders.} (S03)
    \item \textit{They were forced to use these tools that they perceived as creepy and dangerous.} (S14)
    \item \textit{They allow dangerous medical misinformation and animal mistreatment to flourish.} (S29)
    \item \textit{Very dangerous tiktok ‘trends’ like spraying your bathroom mirror with a flammable substance and setting it on fire.} (S29)
\end{itemize}

\subsubsection{Root Causes of Ethical Concerns in Mobile Apps}

\begin{figure}
    \centering
    \includegraphics[width=1\linewidth]{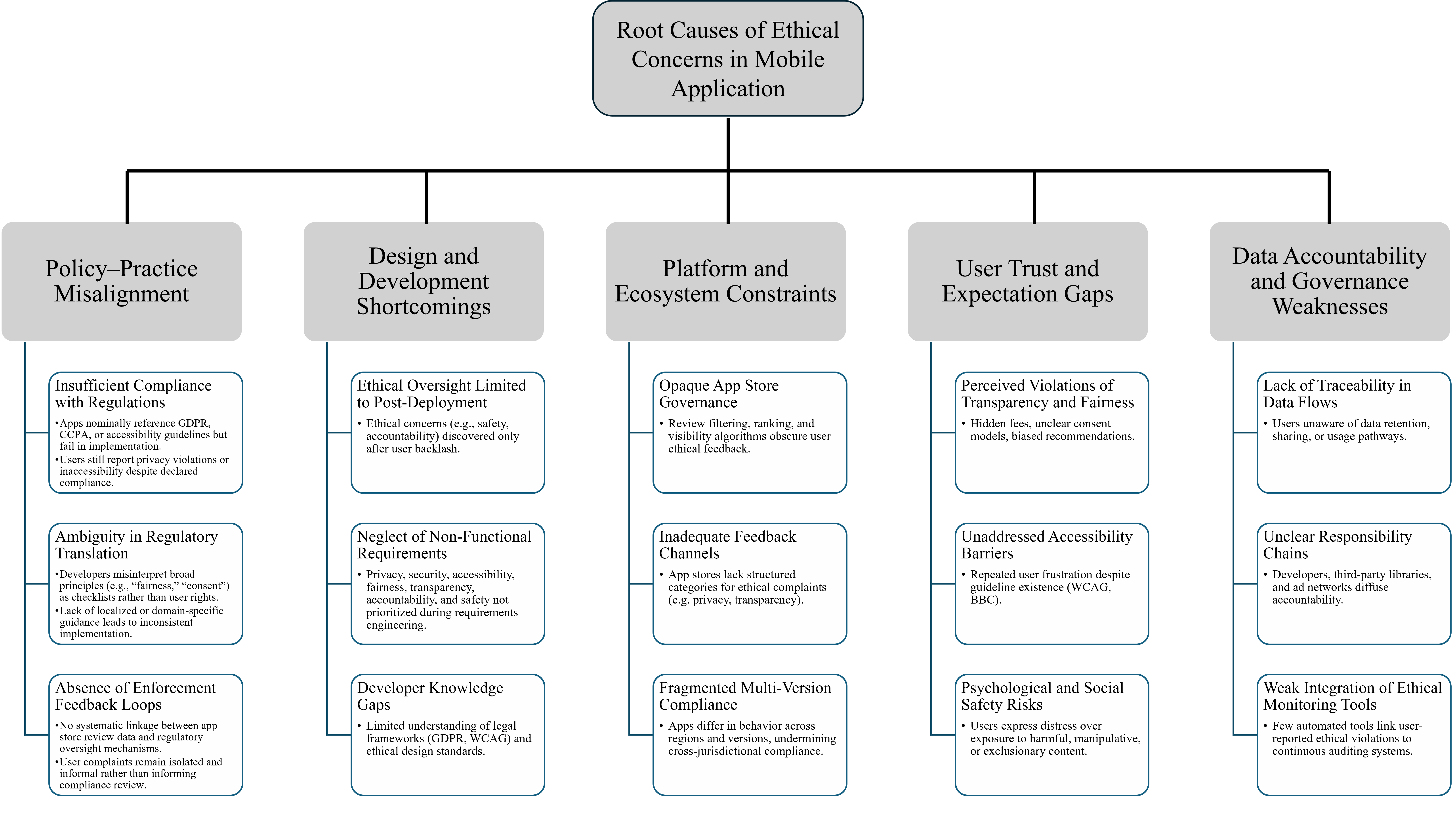}
    \caption{Taxonomy of Root Causes of Ethical Concerns in Mobile Applications.}
    \label{fig:cause_taxo}
\end{figure}

The analysis of the selected studies and user-reported ethical issues led to the identification of several underlying factors that explain why such concerns continue to emerge in mobile applications. These factors were organized into a taxonomy of root causes as shown in Figure \ref{fig:cause_taxo}. The first major cause is \textit{policy–practice misalignment} (S02, S07, S12, S13, S17, S23), where mobile apps often demonstrate superficial compliance with legal or ethical frameworks such as GDPR, CCPA, or accessibility standards, but fail in real implementation. Developers sometimes misinterpret broad principles like consent or fairness as procedural checklists rather than user rights, and the absence of consistent feedback mechanisms between app stores and regulators allows violations to persist without enforcement.

The second root cause lies in \textit{design and development shortcomings} (S03, S04, S05, S16, S18, S30, S36, S37), where ethical oversight is frequently limited to post-deployment stages. Privacy, security, accessibility, and other ethical concerns, commonly regarded as non-functional requirements, are not prioritized early in requirements engineering, and many developers lack adequate understanding of ethical design or regulatory standards. As a result, issues such as data misuse or inaccessible interfaces only surface after public complaints.

Another set of causes stems from \textit{platform and ecosystem constraints} (S10, S20, S21, S27, S28, S33). The governance mechanisms of the app store are largely opaque, and their rating or review algorithms often obscure the ethical feedback of users. Moreover, users have limited avenues to report privacy or accessibility problems in structured ways, and inconsistencies across app versions or regional releases further complicate compliance efforts.

A fourth category relates to \textit{user trust and expectation gaps} (S01, S17, S19, S24, S29). Users frequently perceive violations of transparency and fairness, such as hidden fees, unclear consent flows, and biased recommendation systems. Persistent accessibility barriers also erode confidence, while exposure to manipulative or harmful content introduces psychological and social safety risks.

Finally, \textit{data accountability and governance weaknesses} (S06, S07, S09, S14, S26, S34, S35) play a central role. Many apps lack transparency in data flow and storage, leaving users unaware of how their data is used or shared. Responsibility is often fragmented among developers, third-party services, and advertising networks, making it difficult to trace or assign accountability. Furthermore, there is limited integration of monitoring tools capable of linking user-reported ethical issues with ongoing compliance or auditing systems.

Overall, this taxonomy highlights that ethical concerns reported by users are not isolated incidents, but symptoms of deeper structural, procedural, and governance deficiencies within the mobile app ecosystem. Addressing these root causes requires bridging regulatory intent with developer practice, embedding ethical reflection into design workflows, and improving feedback loops among users, platforms, and policymakers.

\begin{tcolorbox}[breakable,colback=gray!10!white,colframe=gray!75!black,title=Summary of RQ3]
The primary studies revealed a significant number of issues reported by the users in the app reviews. These issues were primarily related to privacy, security, accessibility, transparency, fairness, accountability, and safety. A few examples of such issues include unauthorized data access, account hacks, lack of accessibility features (e.g., screen reader), hidden subscription costs, discrimination based on user demographics, lack of customer support, and harmful app content. We provide a complete list of user-reported issues extracted from the primary studies in \ref{apndxA}.
\end{tcolorbox}

\subsection{RQ4: Future Challenges}
This subsection elaborates on RQ4: What challenges for future research have been identified? We present a structured synthesis of the future work challenges as reported in the primary studies. Through data synthesis, the extracted challenges were categorized into eight major themes, each representing a critical research direction for advancing the analysis and utility of app reviews in exploring users' ethical concerns. By effectively addressing these challenges and capitalizing on the opportunities they offer, future research can achieve substantial advancements in fostering inclusive and reliable mobile apps.\\

\noindent\textbf{App reviews dataset enhancement and diversity}: Many studies highlighted the importance of expanding datasets across platforms (e.g., Android and iOS), languages (especially non-English), and app domains, and enhancing datasets by adding more data labels. These efforts aim to increase representativeness, capture diverse user experiences, and improve the generalizability of the findings.
\begin{itemize}
    \item \textit{Include non-English reviews.} (S01)
    \item \textit{Use general purpose review-authenticity algorithms to validate veracity.} (S02)
    \item \textit{Explore non-english reviews and other less downloaded apps.} (S02)
    \item \textit{Explore gender discussions in other software repositories.} (S03)
    \item \textit{Plan on adding more labels to the dataset, such as bug report or feature request.} (S04)
    \item \textit{Increase the size of our sample.} (S15)
    \item \textit{Include reviews from iOS apps.} (S15)
    \item \textit{Analyze the accessibility update reviews of our dataset to label each review according to the WCAG 2.2 success criteria, which are more specific testable statements that specify requirements for achieving accessibility.} (S16)
    \item \textit{Label our dataset based on the BBC's Mobile Accessibility Guidelines, an accessibility standard more focused on mobile apps and devices.} (S16)
    \item \textit{Considering other applications and more current reviews can enable the discovery of new results, enriching the current results.} (S18)
    \item \textit{Evaluate our approach on more apps from other mobile app stores.} (S22)
    \item \textit{Incorporate data from website users as well.} (S23)
    \item \textit{Include iOS apps and non-English reviews.} (S24)
    \item \textit{Further studies analyzing additional applications from missing domains and with less popularity need to be conducted.} (S29)
    \item \textit{Additional studies should be performed to analyze if the results hold on user feedback written in other app stores and user feedback channels.} (S29)
    \item \textit{Evaluate our approach on larger sets of data and more applications.} (S32)
    \item \textit{Evaluate the proposed approach over other application domains.} (S34)
    \item \textit{Extend our approach to apps in other domains to uncover their strengths and weaknesses.} (S37)
\end{itemize}

\noindent\textbf{App reviews evolution}: This category highlights the importance of investigating the effect of software evolution on users' ethical concerns.
\begin{itemize}
    \item \textit{Investigate whether apps accessibility has increased or decreased over time according to users' perceptions.} (S05)
\end{itemize}

\noindent\textbf{App reviews extraction and classification}: Automated tools and techniques to extract and classify app reviews can significantly enhance the research process, making it more efficient and scalable, which is essential for handling large corpora of app reviews.
\begin{itemize}
    \item \textit{Leverage machine learning models to extract features from the accessibility reviews we identified and found similar reviews in our whole dataset.} (S05)
    \item \textit{Improve the process to select accessibility reviews by increasing our set of keywords.} (S15)
    \item \textit{Optimize the process of extracting and analyzing user reviews using an automatic UUX-Post tool \cite{mendes2017uux}.} (S18)
    \item \textit{Plan on investigating how to automate the multi classification process of accessibility reviews.} (S04)
    \item \textit{Propose a classification for the intention of the accessibility reviews (requests, compliments, promises, threat to uninstall).} (S15)
    \item \textit{Morpho-logical variations in keywords and synonyms can enrich the set of words and make automatic classification more effective.} (S18)
    \item \textit{Planning to explore Active Learning, a well-known machine learning paradigm for classification.} (S20)
    \item \textit{Plan to perform a multi-class classification on the accessibility reviews dividing them into categories such as readability of text, audio, video, UI, gestures, etc.} (S20)
    \item \textit{Only cover the most popular period tracking Android apps in this paper, many other apps and iOS period tracking apps can be analyzed in future work.} (S25)
    \item \textit{Perform sentiment classification using sentence-level lexical-based semantic orientation to evaluate the subjectivity of reviews at the sentence level.} (S26)
    \item \textit{Another improvement would include adding a sentiment analysis to our binary review classifier (SPR/non-SPR). This could help in understanding ambiguous SPR where users complain about requested permissions but still like the application or when users complain but are explicitly fine with a good explanation of the permission usage.} (S28)
\end{itemize}

\noindent\textbf{Content analysis}: Interpreting and modeling ethical concerns-related reviews helps to uncover the patterns in users' concerns and identify the root cause of such concerns, which can help developers to design more user-friendly and inclusive mobile apps.
\begin{itemize}
    \item \textit{Develop new targeted algorithms to identify negative outcomes automatically from reviews or use human-in-the-loop approaches.} (S02)
    \item \textit{Explore differences in healthcare system and regulation policies which are also likely to influence outcomes.} (S02)
    \item \textit{Look into the correlation between the six types of gender discussions and other factors such as user satisfaction, the categories of apps, and the gender of reviewers, through sentiment analysis.} (S03)
    \item \textit{Perform manual content analysis to extract valuable information from our sample.} (S05)
    \item \textit{Leverage topic modeling to automatically identify the main topics addressed by users depending on the declared disabilities or type of interface element mentioned (e.g., font, color, organization).} (S05)
    \item \textit{Explore algorithms other than K-means clustering.} (S07)
    \item \textit{Analyze identified privacy themes in detail.} (S07)
    \item \textit{A deeper exploration into which types of data are more sensitive using sentiment analysis.} (S07)
    \item \textit{Approaches for evaluating and auditing fairness in games and game-like systems to support statistically probable results.} (S09)
    \item \textit{Modules for static and dynamic analysis tools to detect specific values defects.} (S09)
    \item \textit{Improve the process to analyze accessibility reviews by using natural language processing.} (S15)
    \item \textit{Aim to juxtapose the findings from our manual analysis with those generated by employing a Large Language Model (LLM) like ChatGPT-4, assessing its suitability for exploring the intricacies within the context of user review analysis.} (S16)
    \item \textit{Plan to experiment with other clustering techniques and evaluate the extent to which the employed learning models adhere to ethical principles, with a particular focus on fairness.} (S19)
    \item \textit{Use Hark to understand temporal trends in privacy issues, to compare issues based on the emotions dimension, to analyze the type of feedback that leads users to uninstall apps, or to explore particular themes of interest (e.g., ``Blackmailing Concerns'', ``Financial Privacy'', ``Audio Surveillance'', ``Parental Controls'', etc.).} (S21)
    \item \textit{Compare the concern of Airbnb users to the users of ridesharing services such as Uber or Lyft.} (S23)
    \item \textit{More advanced static analysis techniques (i.e., taint analysis) can be applied to analyze the sensitive behavior of the app in future work.} (S25)
    \item \textit{Introduce a comprehensive static and dynamic analysis of mobile payment applications.} (S26)
    \item \textit{Focus on topic modeling techniques such as Latent Dirichlet Allocation (LDA) to discover abstract topics that occur in our collection of documents.} (S26)
    \item \textit{Consider evaluating to which extent users' SPR can be used to create trend analyses of users' attitudes, in particular in response to regulatory (e.g., policies) and system changes (e.g., refactoring of permissions).} (S28)
    \item \textit{Examine additional pattern mining algorithms.} (S32)
    \item \textit{Plan to add additional information such as sentiment, popularity, rating, and emotion to the transaction data.} (S32)
    \item \textit{Expand our research methodology to include a walkthrough approach, which will involve a detailed examination of the interfaces and functionalities of femtech applications. Specifically, we will extract and analyze privacy policies to understand how these apps communicate their data-handling practices to users.} (S35)
\end{itemize}

\noindent\textbf{Ethical concerns guidelines}: Interactive guides and catalogs related to ethical concerns can substantially improve ethical decision-making in the development and evaluation of mobile apps. Additionally, they can assist in refining the methodologies employed for exploring ethical concerns from app reviews.
\begin{itemize}
    \item \textit{Devise an interactive guide in which practitioners can explore accessibility topics and navigate through accessibility reviews to understand the evidence for each accessibility recommendation.} (S05)
    \item \textit{Generate catalogs of comprehensive taxonomies that are indicative of privacy issues in the mobile app market. The generated taxonomies will be used to systematically tune different text summarization, modeling, and classification techniques and identify near optimal configurations (e.g., summary length, similarity thresholds, etc.) to calibrate these techniques.} (S34)
\end{itemize}

\noindent\textbf{User studies}: User perspectives through surveys and interviews can validate the findings from app reviews and provide a more comprehensive understanding of real-world experiences. Additionally, the development of decision-making tools for users can help them understand app behaviors in the context of ethical concerns before installing them.
\begin{itemize}
    \item \textit{Plan to conduct in-situ or naturalistic experiments as part of user studies, along with longitudinal usage analysis. This will allow us to monitor how users interact with these applications in real-world scenarios and enable them to provide real-time feedback.} (S08)
    \item \textit{Development of approaches for generating end-user comprehensible End-User License Agreement (EULA) templates supporting values.} (S09)
    \item \textit{Conduct a more thorough user study to support other categories of risk behaviors.} (S10)
    \item \textit{Conduct user surveys/interviews to explore refusal to use COVID-19 apps.} (S13)
    \item \textit{Interviewing a diverse range of teletherapy app users will allow researchers to better capture their ethical experiences with the technology and gain a more comprehensive understanding of user perceptions and experiences.} (S17)
    \item \textit{Aim to provide a tool that will ease users' understanding towards the privacy behavior of their installed apps, and a reporting tool to support users in a semi-automatic elaboration of app privacy-related reports.} (S27)
    \item \textit{Ranking algorithms could be initialized to mathematically score apps based on invasiveness level of their reviews and their behavior in reality. This way the user can be provided with tailored means to become informed and to make decisions easily, which are, ultimately, important purposes of our system.} (S27)
    \item \textit{Incorporate survey and interview methodologies into our future work. Surveys will enable us to gather quantitative data on a larger scale, providing insights into overarching trends and patterns in user preferences, concerns, and experiences with femtech applications. By designing structured questionnaires, we can systematically collect data on various aspects such as app features, usability, privacy concerns, and satisfaction levels. Through semi-structured interviews with femtech app users, we can explore nuanced aspects of their interactions with the applications, uncovering detailed insights, motivations, and challenges they encounter.} (S35)
    \item \textit{Additional qualitative research using focus groups or interviews to further explore ethical experiences of apps for depression and to advance our preliminary findings in this area.} (S36)
\end{itemize}

\noindent\textbf{Inclusion of stakeholders}: Comprehending the perspectives of various stakeholders in mobile apps, including app owners and developers, can facilitate the formulation of a detailed framework for the design and governance of apps.
\begin{itemize}
    \item \textit{Interviewing therapists or the app companies, which are important parties to consider when offering guidance on the implementation and regulation of teletherapy apps.} (S17)
    \item \textit{Conduct further studies with other stakeholders of AI-based apps such as app owners, developers, and policymakers, and obtain more data from other sources (e.g., GitHub) to generalize our understanding of fairness concerns in AI-based apps.} (S19)
    \item \textit{Conducting a developer survey to ask directly for the incentive of these changes.} (S28)
    \item \textit{Validate the actual usefulness of our approach with software practitioners.} (S32)
    \item \textit{Interviews with stakeholders in the femtech industry, including developers, healthcare professionals, and policymakers, will provide valuable perspectives on the design, implementation, and regulation of femtech solutions.} (S35)
\end{itemize}

\noindent\textbf{Development workflows}: Incorporating ethical concerns assessment tools in the software development lifecycle can support developers in creating inclusive and trustworthy apps from the outset and reduce the need for subsequent analysis and fixes.
\begin{itemize}
    \item \textit{Extend the agile, review-centric methods deployed here to identify and validate early findings about impacts of merging technologies.} (S02)
    \item \textit{Explore and develop practices that can be employed to consider and include gender in mobile app development.} (S03)
    \item \textit{Develop tools that support software developers in locating accessibility bugs and building more accessible apps, as the lack of tools and resources is often cited as a reason for developing inaccessible software.} (S04)
    \item \textit{Development of tools for supporting the inclusion of values throughout the software development lifecycle and the resulting software artefacts, including mobile apps.} (S09)
    \item \textit{Automate accessibility testing during development using the EAAER tool.} (S11)
    \item \textit{Investigate whether developers replied and/or implemented requested features by evaluating different releases of the same app and analysing commits on code repositories.} (S15)
    \item \textit{Manually or automatically evaluate mobile apps to check whether they have accessibility issues that are not reported by user reviews or in issues submitted to code repositories.} (S15)
    \item \textit{Plan to implement a system to automatically complete the whole process in this paper and track the trends of security issues.} (S22)
    \item \textit{Intend to create a system that will automatically perform the whole procedure outlined in this article and monitor security issue trends.} (S33)
    \item \textit{Working prototypes will be developed to assess, through longitudinal studies, the usability and long-term practical significance of our approach. Ultimately, our objective is to help app developers understand how their end-users perceive their app's privacy practices and how these practices can impact their ratings and retention rates.} (S34) 
\end{itemize}

\subsubsection{Research Agenda}
Building on the insights identified across the reviewed studies, we outline key directions for future work aimed at exploring and addressing ethical concerns in mobile app ecosystems. While substantial progress has been made in identifying user-reported issues through app reviews, significant gaps remain in dataset diversity, methodological sophistication, developer tool support, and stakeholder engagement. To guide future research and practice, we recommend four interconnected focus areas: (1) ethical concerns-related app reviews mining, (2) exploring ethical concerns from app reviews using Artificial Intelligence (AI) and advanced natural language processing (NLP) techniques, (3) developing tools to support app developers, and (4) fostering community engagement to ensure inclusive participation. Each direction is grounded in the challenges and opportunities surfaced by this review and aims to advance ethical software practices in a scalable and user-centric manner.\\

\noindent\textbf{Ethical concerns-related app reviews mining}: Future research can focus on improving the diversity and representativeness of app review datasets to effectively address ethical concerns across varied user demographics and platforms. This encompasses the inclusion of non-English reviews, lesser-known apps, and a variety of app categories, particularly from iOS and alternative app marketplaces. Another potential avenue is to broaden labeling methodologies to encompass more granular classifications of ethical concerns derived from related guidelines and frameworks, such as WCAG 2.2 and the BBC Mobile Accessibility Guidelines. Furthermore, our findings indicate that the keyword-based search methodology is predominantly employed for extracting relevant app reviews, which introduces limitations due to its reliance on a predetermined set of related keywords. However, one study (S21) employed the natural language inference (NLI) technique to sample the data, effectively mitigating the limitations associated with keyword-based searches and highlighting the potential of this approach for propelling future research. Moreover, the classification of app reviews into multiple ethical concerns continues to represent a significant challenge for future research.\\

\noindent\textbf{Exploring ethical concerns from app reviews using AI and NLP}: Advancing NLP techniques and AI-driven models presents a promising direction for uncovering patterns and root causes of ethical concerns from app reviews. The primary research has predominantly utilized sentiment analysis and topic modeling approaches for this task and achieved substantial results. However, recent research shows the prevalence of Large Language Models (LLMs) and Retrieval Augmented Graph (RAG) techniques for various requirements engineering tasks \cite{ibtasham2024towards}, which shows the future direction of using these techniques for analyzing ethical concerns from app reviews. Another study shows the potential of utilizing knowledge graphs in this context \cite{sorathiya2024towards}. Additionally, a dual approach combining automated and human-in-the-loop systems can provide scalable yet context-sensitive interpretations of users' ethical concerns.\\

\noindent\textbf{Developing tools to support app developers}: To effectively address the disparity between user-reported issues and implementable enhancements, future research is advisable to prioritize the creation of automated tools that aid developers in the identification, classification, and management of ethical concerns. Such tools may encompass review-focused dashboards, automated systems for categorizing app reviews related to ethical concerns, and plug-ins for integrated development environments (IDEs) that alert developers to ethical risks throughout the development process. Augmenting current tools with functionalities such as real-time sentiment analysis, both static and dynamic assessments for privacy compliance, and checks for accessibility standards can greatly facilitate the ethical design of apps. Subsequent research avenues can establish agile, feedback-oriented workflows that integrate user feedback early within the development cycle, thereby ensuring that ethical responsiveness is incorporated within the software development lifecycle.\\

\noindent\textbf{Encouraging community engagement}: Engaging the extensive stakeholder ecosystem, including end-users, software developers, policymakers, and advocacy organizations, is essential for cultivating ethical mobile ecosystems. Future research can promote systematic user studies, surveys, and interviews to capture a wide array of real-world perspectives, particularly from underrepresented communities. Community-driven reporting and feedback mechanisms, collaboratively developed with end-users, can enhance transparency and foster trust. Similar research by Rui et al. \cite{liu2016privacy} proposed a crowdsourcing platform for designing privacy-aware mobile apps based on user recommendations. Furthermore, interactive platforms that empower users to examine the ethical implications of mobile apps can democratize accountability.

\subsubsection{Replication packages}
16 out of 37 primary studies have made their methodologies and datasets publicly available, which can significantly help future research with the already evaluated datasets and provide baselines for evaluating new approaches aiming at improving app review analysis for exploring ethical concerns. One study (S22) provided the replication package, but it was not accessible, so we have excluded it from our list provided in \ref{apndxB}.

\begin{tcolorbox}[breakable,colback=gray!10!white,colframe=gray!75!black,title=Summary of RQ4]
We identified 78 recommendations from the primary studies for the future work and categorized them into 8 categories, namely, app reviews dataset enhancement and diversity (18 recommendations), app reviews evolution (1 recommendation), app reviews extraction and classification (11 recommendations), content analysis (22 recommendations), ethical concerns guidelines (2 recommendations), user studies (9 recommendations), inclusion of stakeholders (5 recommendations), and development workflows (10 recommendations). Based on these recommendations for future work from primary studies, we provide the research agenda in 4 directions, including ethical concerns-related app reviews mining, exploring ethical concerns from app reviews using AI and NLP, developing tools to support app developers, and encouraging community engagement. We also provide a comprehensive list of replication packages identified from 16 primary studies, which will help to foster future research advancements or replication studies.
\end{tcolorbox}

\section{Discussion} \label{discuss}

\subsection{Discussing RQs}
The results presented in Section \ref{results} show important trends, but also reveal several broader insights about how current research approaches ethical concerns in app reviews. First, we found that many studies focus heavily on privacy and security concerns, while other ethical aspects, such as fairness, accountability, and safety, receive much less attention. For instance, privacy and security appeared in more than half of the 37 studies (Table \ref{tab:concerns}), but fairness and accountability were discussed in fewer than 10. This imbalance highlights a significant gap in the field, as it shows that researchers focus on the most technically measurable ethical concerns while overlooking others equally critical to user trust and inclusion. Future studies could use this gap as an opportunity to broaden the scope of ethical analysis beyond privacy and security, incorporating fairness, accountability, and safety as core elements of ethical app design and evaluation. Another important direction for future research is the need for structured cross-domain analysis of ethical concerns in mobile applications. While existing studies span multiple domains such as mental health, finance, and education, they are typically analyzed in isolation, making it difficult to systematically compare how ethical concerns manifest across domains. A more rigorous cross-domain synthesis could reveal domain-specific patterns, trade-offs, and priorities (e.g., stronger emphasis on privacy and security in healthcare versus other concerns in different domains), thereby providing deeper insights into the contextual nature of ethical issues.

A related gap concerns the empirical relationship between regulatory frameworks and user-reported ethical concerns. While this study references frameworks such as GDPR, CCPA, and accessibility standards (e.g., WCAG) as contextual factors shaping user expectations, the direct influence of regulatory events on observable shifts in app review behaviour remains underexplored. Preliminary evidence from the corpus points toward this connection: Dong et al. (S25) and Nellore et al. (S35) both document measurable changes in privacy-related review sentiment following the overturning of \textit{Roe v. Wade} in 2022, which functioned as a de facto regulatory trigger. Future research should build on this by conducting longitudinal analyses spanning known regulatory milestones, such as the introduction of GDPR enforcement (2018) or the enactment of the CCPA (2020), to assess whether user-reported ethical concerns respond systematically to policy changes. Cross-jurisdictional comparisons across regions with differing regulatory environments (e.g., EU vs. US vs. less-regulated markets) would further clarify how legal frameworks translate, or fail to translate, into developer practice and user awareness.

Second, our analysis shows that many studies rely mainly on quantitative, automated methods such as keyword filtering and machine learning classification. Only a few combined these with qualitative approaches such as manual coding or interviews. For example, studies such as S05 and S17 used thematic analysis to capture ethical concerns in the mental health domain, but most others did not validate their automated findings with human judgment. This limits how deeply researchers can interpret user-reported ethical experiences.

Third, several contradictions appear in the way studies define or interpret ethical concerns. Some studies (e.g., S07, S25, S27) treat privacy as a technical property, focused on data access or permissions, while others (e.g., S01, S17) view it as a matter of user trust and emotional well-being. These different perspectives make it difficult to compare findings across studies. Another recurring issue is the narrow focus on English-language reviews. Only four studies (S11, S12, S13, and S18) analyzed non-English data, meaning that user voices from non-English-speaking regions are largely absent. This limits how representative and inclusive current evidence is. Moreover, several studies used small or convenience samples and did not provide replication packages or code, making reproducibility uncertain. Few papers assessed the reliability or bias of their own data collection and annotation methods, which introduces a potential risk of bias in the reported findings.


Finally, we observed that several studies highlight strong descriptive findings (for example, counts of reviews, app domains, or datasets) but provide little interpretation about why these patterns occur or what they imply for ethical software design. Moreover, although a few papers incorporated perspectives beyond end-users, such as developer interviews, code-level examinations through static or dynamic analyses, or reflections on organizational practices, these insights were often presented only briefly and not integrated into the broader discussion. As a result, potentially valuable contrasts between user experiences, developer intentions, and technical constraints remain underexplored. Future research should therefore focus on synthesizing insights and connecting what users report with methodological or design implications, along with fully engaging multi-stakeholder viewpoints. Bringing developer, organizational, and technical analysis perspectives into dialogue with user-reported data would help move the field beyond cataloguing findings and toward a more holistic understanding of how ethical issues emerge and can be addressed in software ecosystems.

Overall, this discussion shows that research on ethical concerns in app reviews has made progress in mapping the field, but still needs deeper, more balanced, and theory-informed analyses that consider multiple ethical dimensions and diverse user contexts, assess study quality, and address risks of bias to ensure the robustness of future evidence.

\subsection{Implication for software engineers}
When addressing ethical considerations, there are issues that one needs to take into account: (i) ethics is context-dependent, (ii) ethics is not a clear matter of right or wrong, good or bad, or white or black, and (iii) ethical issues are not easy to recognize and deal with \cite{alidoosti2022ethics}. Ethical principles can guide SE, but simply following a set of rules and standards (e.g., the ACM/IEEE Software Engineering Code of Ethics \cite{gotterbarn1997software}) is insufficient because ethical aspects are not easily tested like functional requirements. Therefore, software engineers need explicit and systematic approaches that consider human and non-technical aspects from the early phases of software creation, such as requirements engineering (RE) \cite{biable2023proposed, hosseini2016foundations}.

RE is a critical stage in the software development life cycle (SDLC), as its success directly impacts the success of software projects \cite{shah2014review, mudduluru2016value}. Even if a software system is implemented correctly, it will fail if it does not meet the necessary requirements \cite{mudduluru2016value}. RE is a crucial yet challenging phase of SDLC \cite{shah2014review, mudduluru2016value} as it determines exactly what needs to be developed and ensures the specification and validation of detailed technical and non-technical requirements \cite{mudduluru2016value}. Additionally, this stage is particularly challenging when it comes to addressing ethical concerns, which are non-technical attributes of software. In reality, ethical considerations are often overlooked during RE \cite{shah2014review}. This is because software engineers usually act in isolation from the software users \cite{begier2010users} and are generally culturally unfamiliar with some or most of their users \cite{Vallor2013AnIT}. Consequently, our paper bridges this gap by surveying the relevant literature and providing actionable insights to software developers and practitioners.

In RQ3, we extracted the user-reported issues from the primary studies related to the software quality attributes (ethical concerns): privacy, security, accessibility, transparency, fairness, accountability, and safety. These issues provide first-hand experience of users and help developers understand how their apps are perceived from an ethical viewpoint. These issues also provide directions for further app enhancements to implement non-functional requirements and address users' ethical concerns. Furthermore, we also provide a seven-stage analysis pipeline, derived from primary studies, that can guide software companies to build automated frameworks to elicit non-functional requirements from user feedback.

\subsection{Research agenda for requirements engineering}
In RQ4, we provided a research agenda in four directions based on the future work challenges identified from the primary studies. These future avenues are ethical concerns-related app reviews mining, exploring ethical concerns from app reviews using AI and NLP, developing tools to support app developers, and encouraging community engagement. These avenues also provide directions for data-driven requirements engineering, leveraging app reviews as large-scale user feedback to improve software quality attributes. For instance, our research agenda proposes to develop automated frameworks and IDE plug-ins for developers to incorporate ethical standards and address users' ethical concerns in the early phases of the app development process. Furthermore, mining ethical concerns-related app reviews can significantly help in identifying and modeling users' goals following goal-oriented requirements engineering. For instance, Tushev et al. \cite{tushev2020digital} suggested a Feature-Goal Interdependency Graph to address safety and accountability concerns in the sharing economy domain by identifying user-suggested features from Twitter data. Ultimately, our goal is to foster a more responsible and user-centered development process.

\begin{table}[h]
    \centering
    \caption{Taxonomy: Integrating Ethical Concerns Across SE Life Cycle Stages.}
    \label{tab:process_taxo}
    \begin{tabular}{>{\raggedright}p{2.2cm}p{13cm}}
        \textbf{SE Process} & \textbf{Integration of Ethical Concerns} \\
        \hline
        \hline
        Requirements Engineering (S03, S07, S09, S10, S21, S25, S28, S29, S34) & Identify and translate ethical concerns reported by users (e.g., privacy, fairness, accessibility) into explicit functional and non-functional requirements. These concerns should be documented as measurable requirements, ensuring that user expectations, legal standards (e.g., GDPR, WCAG), and domain-specific ethical guidelines are addressed from the beginning.\\
        Design (S04, S11, S13, S14, S30, S36) & Incorporate *Value-Sensitive Design*, *Privacy by Design*, and *Inclusive Design* principles to ensure that user autonomy, transparency, and inclusivity are embedded into the interface and interaction design. Avoid manipulative design patterns (“dark patterns”) that undermine user consent or fairness.\\
        Implementation (S06, S08, S22, S26, S32, S33) & Enforce ethical coding practices by applying secure development techniques, data minimization principles, and transparent data-handling methods. Developers should use modular code structures that facilitate traceability of ethical functions such as consent tracking, privacy settings, or accessibility support.\\
        Testing \& Quality Assurance (QA)  (S05, S19, S20, S24) & Develop and execute ethical test cases that assess compliance with ethical requirements. This includes testing for accessibility (e.g., WCAG compliance), fairness (e.g., unbiased algorithmic outcomes), security vulnerabilities, and privacy protection. Testing also involves simulated user interactions to identify potential ethical risks before release.\\
        Deployment (S11, S12, S23, S31) & Ensure that deployed systems follow ethical defaults—such as opt-in data sharing, explicit consent prompts, and clear communication of updates or data use changes. Deployment documentation should include ethical compliance statements and change logs related to user privacy or accessibility features.\\
        Maintenance \& Monitoring (S01, S07, S15, S16, S20, S27, S36, S37) & Continuously monitor user feedback and app reviews to identify emerging ethical issues such as privacy violations, safety concerns, or accessibility failures. Integrate automated tools and review-mining systems that can flag ethical complaints and trigger corrective updates, ensuring sustained ethical compliance after release.\\
        Project Management (S02, S09, S14, S37) & Integrate ethical risk assessment and compliance tracking into project management activities. Regular stakeholder engagement, including user representatives, legal experts, and accessibility advocates, ensures that ethical considerations remain central throughout the project life cycle. Management frameworks (e.g., IEEE, ISO standards) can support accountability by embedding measurable ethical performance indicators.\\
        \hline
    \end{tabular}
\end{table}

\subsection{Integrating Ethical Concerns Across Software Engineering Life Cycle Stages}
In earlier sections, we primarily emphasized ethical concerns as non-functional requirements within the RE phase; however, it is evident that addressing ethics effectively requires continuous attention throughout the entire SE process \cite{alidoosti2022ethics, gotterbarn1997software, mitchell2022incorporating, Vallor2013AnIT}. Based on the synthesis of findings from the reviewed studies and established software ethics frameworks, we developed a taxonomy that connects ethical concerns with the major stages of the software development life cycle, as shown in Table \ref{tab:process_taxo}. This taxonomy highlights how privacy, security, accessibility, fairness, transparency, accountability, and safety can be systematically integrated into each phase of the development process.

This taxonomy underscores that ethics in software engineering is not a single-step consideration but a continuous, iterative process. During requirements engineering, user feedback from app reviews and other channels should be analyzed to extract ethical expectations and transform them into verifiable requirements. In the design stage, these expectations evolve into design principles that promote user trust, autonomy, and inclusivity. The implementation phase ensures that ethical intent translates into practice through secure and responsible coding. In testing, ethical compliance is verified alongside traditional quality attributes, enabling teams to detect and mitigate biases, accessibility failures, or privacy breaches. Following deployment, the maintenance and monitoring phase plays a crucial role in ensuring that ethical integrity is preserved over time. As users continue to report new issues, developers can iteratively refine and adapt ethical features. Finally, project management acts as the backbone of this process, embedding accountability, documentation, and cross-functional collaboration to ensure that ethics are systematically integrated into every development decision.

This taxonomy extends the traditional view of ethical concerns as non-functional requirements by embedding them throughout the full development life cycle. Such an approach encourages continuous ethical reflection and ensures that mobile applications remain compliant, inclusive, and trustworthy in real-world contexts.

\subsection{Broader Implications}
This survey identified four key directions for future research based on the analysis of 37 studies. These directions include improving how ethical concern-related app reviews are mined, using advanced AI and NLP techniques to analyze user feedback, developing tools that help developers address ethical issues early in the app development process, and fostering community engagement through user studies and stakeholder collaboration. These areas aim to make mobile apps more inclusive, trustworthy, and aligned with user expectations.

Beyond SE, the findings of this survey have broader implications for other research and practice communities. For instance, HCI researchers can use the accessibility issues identified in app reviews to improve interface design for diverse user groups. Ethics scholars may find real-world examples of how principles like fairness and accountability are experienced by users. Policy makers can draw on insights from app reviews to inform digital regulations, especially around privacy and safety. App designers and product managers can use this knowledge to create apps that are not only functional but also ethically sound and user-friendly. We encourage a more holistic and inclusive approach to mobile app development and evaluation by summarizing the research agenda and highlighting its relevance to a wider audience.


\section{Threats to Validity} \label{threats}
Assessing threats to validity is crucial for ensuring the scientific value of literature surveys in SE \cite{zhou2016map}. This section presents the threats to validity that may have affected the results presented in our survey. The categories of threats to validity defined by Zhou et al. \cite{zhou2016map} have been analyzed:

\subsection{Construct validity}
Construct validity refers mainly to the review protocol and the degree of inclusion of the selected search string, electronic databases, and selection criteria concerning the research questions. To minimize the threats to construct validity, a validation procedure was included for each step of the review protocol. For instance, the search string was defined as generic as possible, not to exclude any potentially relevant studies following PICO guidelines. To mitigate potential bias from our choice of databases and to ensure no studies were overlooked, we employed one-step backward and forward snowball sampling. Despite our efforts, the possibility remains that some key studies might have been overlooked. Furthermore, we acknowledge that the exclusion of publication-specific venues, such as Springer and Elsevier, might have excluded some relevant studies.

In addition, only English-language, peer-reviewed papers were included, which introduces a language bias and may exclude research conducted in other regions or languages. Future updates of this review could benefit from including non-English publications to achieve a more global view of ethical concerns in app reviews. Our search query focused on seven ethical concerns that are well established in the literature. This means that other possible or emerging ethical issues, such as sustainability or explainability, may not have been captured in our search. As a result, the review may be slightly biased toward these main categories. However, given the wide range of possible ethical topics, it was necessary to focus on the most common ones to keep the study clear and practical.

\subsection{Internal Validity}
Internal validity refers to the reliability of the cause-and-effect relationship between the results and the methodology of the survey. In this survey, the search procedure was mainly performed by a single researcher, so the results may be subject to selection bias. To reduce this, we defined clear inclusion/exclusion criteria, and a validation procedure was performed in which the second author reviewed a randomly selected set of studies, such that each study was reviewed by both authors.

\subsection{External Validity}
External validity refers to how the results of the survey apply to the subject area, which, in this case, is exploring users' ethical concerns from app reviews. The findings may be influenced by the rapidly evolving nature of mobile technology and ethical considerations in SE. We specified the period covered in our survey and discussed the impact of recent developments. We derived seven common ethical concerns and categorized the primary studies into these based on their purpose of analysis and the user-reported issues explored in them. We used established guidelines from the literature to derive these ethical concerns and mitigate the publication bias; however, we acknowledge that this categorization may impact the generalizability of our survey.

Further, this review focused solely on published academic studies, excluding grey literature, preprints, industry reports, and theses. While this ensured a consistent level of peer-review quality, it may have omitted valuable insights from practitioners or emerging findings that have not yet appeared in formal venues. Such publication bias may also favor studies with positive or confirmatory results. Incorporating grey literature into future reviews could provide a more balanced picture of current practices. Additionally, our review specifically targeted studies that analyzed app reviews to explore users’ ethical concerns. Other relevant data sources, such as developer forums, issue trackers, or social media platforms, were intentionally excluded to preserve focus. Although this improves consistency, it limits the scope of our conclusions. Similarly, the dominance of studies using English-only review datasets restricts cultural diversity and may overlook ethical concerns expressed in other languages or contexts.

\subsection{Conclusion validity}
Conclusion validity pertains to the replicability of the findings. To accomplish this objective, we established a comprehensive protocol for data extraction and formatting. This protocol was formulated by one author and subsequently evaluated by their supervisor. We have made the protocol and extracted data accessible for future replication or amendments to our survey. Furthermore, we automated each step wherever feasible. However, the inherent subjectivity involved in certain aspects of the survey inhibits our ability to assure the identical results should other researchers undertake the literature survey; however, we believe that the overall results should be similar.

\section{Conclusion and Future Work} \label{conclude}
In this comprehensive literature survey, we present our findings on analyzing app reviews to explore users' ethical concerns. Through a systematic search, we identified 37 relevant studies published between 2012 and early 2025, which we thoroughly examined to answer our research questions. The findings have revealed a growing interest in the research area, with most publications appearing in Human-Computer Interaction conferences and journals, such as ACM Human-Computer Interaction and the CHI Conference on Human Factors in Computing Systems. The research in this area is likely to continue gaining importance as a consequence of increased interest in mobile app development amid rising awareness about ethics and ethical concerns among both the mobile app development industry and app users.

Our survey offers essential insights and highlights the methodologies employed by researchers in the examination of app reviews. A variety of techniques for collecting and analyzing app reviews were recognized, encompassing automatic, manual, and hybrid approaches, as well as the use of pre-existing datasets. Despite advancements in ethical guidelines and frameworks, several ethical concerns-related issues were identified in app reviews, including unauthorized data access, lack of data encryption, poor interface designs, hidden and unfair costs, inadequate customer support, and harmful communities. In this literature survey, we also identify several future challenges, categorized into eight major themes, including the expansion, extraction, and classification of app reviews. 

Our findings provide a research agenda in four key directions, including ethical concerns-related app reviews mining, exploring ethical concerns from app reviews using AI and advanced NLP techniques, developing tools to support app developers, and fostering community engagement to ensure inclusive participation. The use of Covidence software significantly enhanced our survey by improving both organizational and operational efficiency. We addressed areas where it was inadequate by employing a spreadsheet to facilitate better data extraction and visualization. Furthermore, one-step forward and backward snowballing was advantageous, resulting in the inclusion of three additional studies in our survey.

In conclusion, we believe that this survey can help communicate knowledge on analyzing app reviews to explore and address users' ethical concerns. We are confident that our effort will inspire scholars to advance the research area and assist them in positioning their new works aptly.

In the future, we aim to broaden the scope of our investigation by incorporating studies that examine user feedback from platforms such as Reddit, Twitter, or E-Commerce Websites. This approach will yield a more comprehensive understanding of user perspectives, not only regarding mobile apps but also encompassing any software category, including websites, desktop apps, or web extensions.

\section{CRediT authorship contribution statement}
\textbf{Aakash Sorathiya}: Writing – review \& editing, Writing – original draft, Visualization, Validation, Methodology, Investigation, Formal analysis, Data curation, Conceptualization. \textbf{Gouri Ginde}: Writing – review \& editing, Validation, Supervision, Project administration, Funding acquisition, Conceptualization.

\section{Declaration of competing interest}
The authors declare that they have no known competing financial interests or personal relationships that could have appeared to influence the work reported in this paper.

\section{Data availability}
The link to the data (\url{https://bit.ly/3GiTKli}) is specified in the paper.

\section{Acknowledgement} \label{ack}
This research was partially supported by the Natural Sciences and Engineering Research Council of Canada, NSERC Discovery Grant RGPIN-2023-03365.


%
%

\bibliographystyle{elsarticle-num-names}
\bibliography{ref}

\appendix

\section{Objectives of Primary Studies} \label{apndxAA}

\textit{S01: To analyze users' real-life experiences and concerns regarding mobile mental health apps, with a focus on usability, efficiency, and ethical concerns.}

\textit{S02: To explore negative outcomes of freemium models in mental health apps for vulnerable populations.}

\textit{S03: To unpack gender discussions about apps from the user perspective by analyzing app reviews to support the development of gender-inclusive apps.}

\textit{S04: To understand how accessibility issues are expressed in user reviews and map them to technical accessibility guidelines (WCAG 2.1).}

\textit{S05: To analyze accessibility reviews associated with visual disabilities or eye conditions, identifying barriers faced by users and their impact on app usability.}

\textit{S06: To analyze security issues in mobile health and medical apps, including user perceptions of security through app reviews.}

\textit{S07: To analyze user privacy concerns at scale using natural language processing (NLP).}

\textit{S08: To evaluate security, privacy, user interaction, and accessibility in mobile payment apps and correlate user perceptions (from app reviews) with actual vulnerabilities detected by tools.}

\textit{S09: To detect and categorize violations of the human value of honesty in mobile apps and understand developers' perspectives on honesty violations.}

\textit{S10: To automatically infer security and privacy-related behaviors of Android apps from user reviews, focusing on the ``review-to-behavior fidelity'' - how well user reviews reflect actual app behaviors.}

\textit{S11: To identify accessibility issues that prevent users from downloading or using emerging apps, with a case study on COVID-19 apps.}

\textit{S12: To identify and characterize undesired app behaviors from app reviews that violate app market policies.}

\textit{S13: To compare and understand user perceptions about privacy concerns between COVID-19 and social media/productivity apps.}

\textit{S14: To understand how end-users perceive and experience privacy, security, and safety risks in education technology.}

\textit{S15: To understand the impact of accessibility complaints on app ratings and determine if user reviews could serve as a driver for improving the mobile app accessibility.}

\textit{S16: To understand the impact of app updates on accessibility from the users' perspective and identify accessibility issues introduced by app updates.}

\textit{S17: To explore ethical experiences of users with teletherapy apps from app reviews.}

\textit{S18: To identify accessibility problems in educational apps for autistic users and evaluate compliance with accessibility guidelines.}

\textit{S19: To investigate fairness concerns in AI-based mobile apps from the users' perspective using automated classification and summarization techniques.}

\textit{S20: To automatically identify accessibility-related user reviews.}

\textit{S21: To discover and summarize privacy-related feedback at scale.}

\textit{S22: To identify and summarize security issues in mobile apps based on user reviews.}

\textit{S23: To identify users' concerns and challenges in the lodging sharing economy, with a case study of the Airbnb app.}

\textit{S24: To analyze sentiments in accessibility-related user reviews to understand how users feel about the accessibility features of mobile apps.}

\textit{S25: To understand user concerns about privacy issues in period tracking apps, particularly after the overturning of Roe v. Wade.}

\textit{S26: To evaluate privacy and security concerns expressed by users in mobile payment apps.}

\textit{S27: To identify privacy and security concerns in mobile apps by performing automated summarization of user reviews.}

\textit{S28: To measure the impact of user reviews on app security and privacy: how often users raise security/privacy concerns in reviews; how developers respond to such reviews; whether these reviews lead to measurable updates in app security/privacy}.

\textit{S29: To develop automated approaches to classify app reviews into different types of ethical concerns.}

\textit{S30: To assess accessibility compliance of the Blackboard mobile app for users with disabilities.}

\textit{S31: To identify user concerns in Metaverse games using text mining approaches.}

\textit{S32: To understand the underlying context of users' ethical concerns.}

\textit{S33: To develop a comment summarization framework that provides a detailed summarization of the app’s security issues.}

\textit{S34: To perform unsupervised summarization of privacy concerns using domain-specific keywords.}

\textit{S35: To explore user perspectives on accessibility, interface design, and privacy of female healthcare apps and analyze how user perceptions of these apps have changed over time.}

\textit{S36: To understand the ethical experiences of users with apps for depression.}

\textit{S37: To evaluate mental health apps based on user reviews and identify issues related to usability, privacy, billing, and customer support.}

\section{User-reported Ethical Concerns-related Issues} \label{apndxA}

\textbf{Privacy-related issues extracted by the primary studies from app reviews. }

\begin{itemize}
    \item \textit{The user was terrified since their encounter with the app seemed like they were being monitored.} (S01)
    \item \textit{Collecting superfluous information from patients in an insensitive and inaccurate manner.} (S01)
    \item \textit{Contained very little or no information about when and how the collected data is disseminated or shared with other stakeholders or third party.} (S01)
    \item \textit{Excessive permission requests.} (S06)
    \item \textit{Reviews that make comments about the app asking for too many permissions.} (S07)
    \item \textit{Location permission is the most discussed permission, appearing in over 40,000 reviews.} (S07)
    \item \textit{Users complain that too much personally identifiable information (PII) is being collected.} (S07)
    \item \textit{Users indicate their suspicion i.e., there is no good reason for the data collection.} (S07)
    \item \textit{Frustration with privacy controls.} (S07)
    \item \textit{We saw tracking location, purchase history, contacts, and other personal information as a dominant concern.} (S07)
    \item \textit{Users’ have previously expressed concerns with service providers selling their data.} (S07)
    \item \textit{User has difficulty accessing their account.} (S08)
    \item \textit{User’s account was closed without proper in-depth investigation.} (S08)
    \item \textit{User was not provided with an alternative layer of authentication, and got locked out of their account.} (S08)
    \item \textit{Inability to remove personal data.} (S08)
    \item \textit{Unnecessary data collection.} (S08)
    \item \textit{Perceived unnecessary permissions requested.} (S08)
    \item \textit{Data being shared with third-parties.} (S08)
    \item \textit{Inadvertent keeping of user’s information.} (S08)
    \item \textit{Over-privileged permission.} (S10)
    \item \textit{User reviews showed that many mobile users have privacy concerns or misconceptions about new emerging apps such as COVID-19 contact tracing apps} (S11)
    \item \textit{Privacy leak.} (S12)
    \item \textit{Payment deception.} (S12)
    \item \textit{User data collected out of the scope of the application.} (S13)
    \item \textit{The information they collect to promote their products or services.} (S13)
    \item \textit{Collect user’s current address and places they regularly visit, businesses they use, and people they are close by.} (S13)
    \item \textit{User information is used in research and some users are not aware that their data is used in this way.} (S13)
    \item \textit{Try to collect as much data as possible, whether or not the user knows about it.} (S13)
    \item \textit{Collection of data, such as browsing history, device information, authentication tokens (such as photo ID shown before taking exams), and financial information (e.g., credit card number used to pay for the service remotely).} (S14)
    \item \textit{Collecting granular data through location and behavioral tracking.} (S14)
    \item \textit{Spying on spouse.} (S21)
    \item \textit{Excessive permissions.} (S21)
    \item \textit{Unneeded contacts access.} (S21)
    \item \textit{Unneeded location access.} (S21)
    \item \textit{Data stealing.} (S21)
    \item \textit{Keeps secret.} (S21)
    \item \textit{Personal info collection.} (S21)
    \item \textit{It wants to track and report the sites you visit, and which apps you use—which seems unnecessary and somewhat creepy.} (S21)
    \item \textit{Requested too many permissions that don’t seem to be needed.} (S22)
    \item \textit{They steal contact lists and information.} (S22)
    \item \textit{Access to all of your personal information.} (S23)
    \item \textit{No trust in any period tracking app that uploads user data to the server.} (S25)
    \item \textit{Apps disclosing personal data due to law enforcement.} (S25)
    \item \textit{Providing personal information when creating an account raises privacy issues, and logging in with a third-party account may release their data to an undesired third-party company.} (S25)
    \item \textit{App does not provide an alternative layer of authentication.} (S26)
    \item \textit{Issues in requesting a refund of their money when they are unable to access their account.} (S26)
    \item \textit{Concerns with the amount of info required for verification as well as the point at which certain verification is initiated by the app developers.} (S26)
    \item \textit{Inability to remove personal data from the app.} (S26)
    \item \textit{Revealing personal information to third-parties.} (S26)
    \item \textit{Request access for permission to access mobile data or functionality.} (S26)
    \item \textit{Unauthorized data charges.} (S27)
    \item \textit{Unintended data disclosure.} (S27)
    \item \textit{Access calendar and confidential information.} (S29)
    \item \textit{Cannot log in to their accounts and that new account creation requests are answered late.} (S31)
    \item \textit{Inability to access and control their content.} (S31)
    \item \textit{Collects SSN and bank login info.} (S34)
    \item \textit{They sold info and social security number.} (S34)
    \item \textit{Wants too much personal information, unrelated to its purpose.} (S34)
    \item \textit{Requesting email/Facebook data.} (S34)
    \item \textit{Asking for sensitive information before the free trial.} (S34)
    \item \textit{Collecting patient information (health info, job history, lastname).} (S34)
    \item \textit{Requesting unnecessary permissions (location, device data).}
    \item \textit{Sharing and selling users’ data.} (S34)
    \item \textit{Requesting access to camera and microphone.} (S34)
    \item \textit{Collecting identification information (SSN, birthdate, driver’s license).} (S34)
    \item \textit{Collecting Financial information (bank statements, income, tax info, login info).} (S34)
    \item \textit{Selling personal data and exploiting a vulnerable process, and exploiting women.} (S35)
    \item \textit{Unauthorized access to accounts and loss of personal data.} (S35)
    \item \textit{Collection of sensitive user data before confirming access to the intervention.} (S36)
    \item \textit{Lack privacy policies or fail to provide clear and concise information on data collection, sharing, and use.} (S37)
    \item \textit{Full telemetry milking.} (S37)
    \item \textit{Privacy policy is unacceptable.} (S37)
\end{itemize}

\textbf{Security-related issues extracted by the primary studies from app reviews.}

\begin{itemize}
    \item \textit{Reviews reported intrusive permissions or sensitive permission requests.} (S06)
    \item \textit{Explicit mentions of malware or suspicious activity.} (S06)
    \item \textit{Review explicitly mentioned the lack of secure HTTPS.} (S06)
    \item \textit{Virus} (S12)
    \item \textit{Illegal background behavior (e.g., SMS).} (S12)
    \item \textit{Excessive network traffic.} (S12)
    \item \textit{Hidden app.} (S12)
    \item \textit{Illegal redirection.} (S12)
    \item \textit{Permission abuse.} (S12)
    \item \textit{Illegitimate update.} (S12)
    \item \textit{Intrusive access to the device and peripherals.} (S14)
    \item \textit{Surveillance and control.} (S14)
    \item \textit{Invasive, parasitic malware.} (S22)
    \item \textit{Added viruses to phone.} (S22)
    \item \textit{App overcharged automatically without permission.} (S22)
    \item \textit{Hacked bank account.} (S22)
    \item \textit{Changed system settings automatically by itself.} (S22)
    \item \textit{Completely unnecessary bloatware.} (S22)
    \item \textit{Store user information on the server-side.} (S25)
    \item \textit{Difficult or impossible to revert wrong transactions.} (S26)
    \item \textit{Tracking and spyware.} (S27)
    \item \textit{Phishing.} (S27)
    \item \textit{Malwares and computer viruses.} (S31)
    \item \textit{Authentication and identity theft.} (S31)
    \item \textit{Intellectual property theft.} (S31)
    \item \textit{Account getting hacked and all of money stolen twice.} (S31)
    \item \textit{Private chat is accessed by 3rd party without permission.} (S34)
    \item \textit{Patient, doctor confidentiality breached.} (S34)
    \item \textit{They sell screen recordings of customers to third party.} (S34)
    \item \textit{Absence or inadequacy of data encryption measures.} (S35)
\end{itemize}

\textbf{Accessibility-related issues extracted by the primary studies from app reviews. }

\begin{itemize}
    \item \textit{Readability of text, change font size, reflow.} (S04)
    \item \textit{Volume too high, audio plays automatically, ability to pause audio.} (S04)
    \item \textit{Subtitles, video plays automatically, zooming into, navigate video.} (S04)
    \item \textit{Color scheme, elements’ (buttons, links, etc.) size.} (S04)
    \item \textit{Controls, e.g., volume control, enough time to interact with controls.} (S04)
    \item \textit{No need to scroll or memorize information.} (S04)
    \item \textit{Hand movement, perform action with a gesture.} (S04)
    \item \textit{Indicating and correcting wrong input.} (S04)
    \item \textit{Haptic feedback.} (S04)
    \item \textit{Reviews on the necessity of increasing font size, as well as updating the apps for zoom features.} (S05)
    \item \textit{Reviews that request dark mode features mainly to avoid eyestrain because they use the app at night.} (S05)
    \item \textit{Reviews associated with keyboards.} (S05)
    \item \textit{Reviews related to color blindness, in which users complain they cannot differentiate between traffic and read-confirmation color encodings.} (S05)
    \item \textit{Difficult for users to navigate, locate desired features, and complete transactions efficiently.} (S08)
    \item \textit{They could not download or install these apps since they do not support their iOS or Android version.} (S11)
    \item \textit{Apps are only accessible by users using a local country app store and they are not available in the international store.} (S11)
    \item \textit{Apps were not accessible by non-residents since they require a national ID of some form.} (S11)
    \item \textit{Only supporting one or two languages.} (S11)
    \item \textit{Their kids were unable to use COVID-19 apps.} (S11)
    \item \textit{Using Bluetooth for tracing, which affects the connectivity with other devices such as headphones and smartwatches.} (S13)
    \item \textit{Users mentioned issues with theme/mode, zoom, customization, media, font, contrast, impairment, flickering, and language.} (S15)
    \item \textit{Numerous issues related to the color scheme of the user interface.} (S16)
    \item \textit{Removed or failed to include dark mode, night mode, or dark theme options.} (S16)
    \item \textit{Challenges in navigating apps, finding options, reading images, and encountering compatibility issues with screen readers.} (S16)
    \item \textit{Lack of customization options and the inability to adjust font size.} (S16)
    \item \textit{Lack of accessibility settings.} (S16)
    \item \textit{Text being too small (similar to Font), difficulties using voice-to-text features, removal of text reflow or word wrap functionality.} (S16)
    \item \textit{Buttons lacking labels necessary for assistive technology.} (S16)
    \item \textit{App is not delivering a clear message about the error preventing its use.} (S18)
    \item \textit{Font could not be changed.} (S18)
    \item \textit{Impossible to change the character’s speech.} (S18)
    \item \textit{Not providing adequate responses for correct actions or alerts.} (S18)
    \item \textit{Makes an error when demonstrating word pronunciation.} (S18)
    \item \textit{Users reported issues with dark mode, zoom, volume settings, font size, background color, voice command, text-to-speech, eyestrain, screen reader, language, and navigability.} (S20)
    \item \textit{Unclear error messages, speed, glitches and misplaced elements became a hindrance when completing transactions.} (S26)
    \item \textit{Buttons or functionalities hard to detect.} (S26)
    \item \textit{Reviews requesting a rotating screen and captions for aid with disabilities.} (S29)
    \item \textit{Describe a need for captioning.} (S29)
    \item \textit{The application did not move to the next question but went and submitted the exam test.} (S30)
    \item \textit{Add dark mode, plus the feature to customize, organize and categorize modules.} (S30)
    \item \textit{Hard to navigate and often causes glitches in test submission.} (S30)
    \item \textit{Lower levels of playfulness in games.} (S31)
    \item \textit{Disconnecting constantly in the middle of games.} (S31)
    \item \textit{Notes and Pop-Ups inaccessible.} (S35)
    \item \textit{Poor responsiveness.} (S36)
    \item \textit{Interrupting phone calls, sounds and volume settings, or battery life and memory.} (S36)
    \item \textit{Poor layout of interface elements, bad colour scheme, graphics and text rendering issues, and choppy animations of mental health apps.} (S37)
    \item \textit{Difficulty accessing desired features or locations within the app.} (S37)
\end{itemize}

\textbf{Transparency-related issues extracted by the primary studies from app reviews. }

\begin{itemize}
    \item \textit{Users have also expressed their dissatisfaction with the sometimes-confusing nature of these standard guidelines.} (S01)
    \item \textit{Some apps provide unclear and incomplete instructions in the interface for searching for therapists.} (S01)
    \item \textit{The algorithms utilized in the apps to match therapists in regards to their needs were ineffective.} (S01)
    \item \textit{It often took a lot of time and effort on the user end to understand that a particular functionality in the app that attracted her is not available in the free version.} (S01)
    \item \textit{The payment process’ transparency was also challenged by users.} (S01)
    \item \textit{Some apps were vague in how they described the MH conditions that they targeted, with the consequence that users may download apps that do not actually help their condition.} (S02)
    \item \textit{Users also felt that payment plan descriptions were imprecise, leaving them unclear about which features were offered in the free version, and confused about the charging model.} (S02)
    \item \textit{Withdrawing from a free trial also confusing.} (S02)
    \item \textit{User talks about how the app explicitly encouraged them to engage by establishing initial treatment goals like dangling a carrot, but without revealing that subscriptions are needed to unlock these benefits.} (S02)
    \item \textit{Users found it inaccurate or less useful when the translation feature does not provide any clue of gender.} (S03)
    \item \textit{Users perceive the cancellation and refund policy as unfair, nontransparent, or deliberately misleading.} (S09)
    \item \textit{User perceives that the advertised features and functionalities of the app as described by the developers are not contained in the app.} (S09)
    \item \textit{Unfair or nontransparent automatic subscription processes.} (S09)
    \item \textit{Inconsistency between functionality and description.} (S12)
    \item \textit{Frustrated as the cost was not disclosed until they finished the sign-up process, wishing pricing information was clear up front.} (S17)
    \item \textit{Their employed algorithms unfairly and non-transparently treat user-generated content.} (S19)
    \item \textit{Same places show at different prices for the same dates.} (S23)
    \item \textit{Listing where I am currently staying look different depending on if you are on the App or the website. On the App version it says nothing about paying for people more than 3, paying a cleaning fee, or needing a security deposit that are stated on the Website. The cost per day is also different.} (S23)
    \item \textit{User questions new permissions needed by the app.} (S29)
    \item \textit{Users describing unclear rules regarding bans and deceitfulness concerning data collection.} (S29)
    \item \textit{Attempts to access their device’s camera, calendar, and contacts without explaining what will be done with this access.} (S29)
    \item \textit{Less transparent costing.} (S36)
    \item \textit{Insufficiency of information regarding app costs and billing practices.} (S36)
    \item \textit{It is completely untransparent what happens with the emotional data of end-users.} (S37)
\end{itemize}

\textbf{Fairness-related issues extracted by the primary studies from app reviews. }

\begin{itemize}
    \item \textit{Some users have expressed their dissatisfaction with the moderation policies, claiming that they are too strict to express their negative emotions or to vent.} (S01)
    \item \textit{Users often complained how the moderation is ``two-faced''.} (S01)
    \item \textit{Users have occasionally expressed their disappointments with moderators ``feeling entitled'' and ``abusing their power'' by showing conscious biased behavior.} (S01)
    \item \textit{Moderators also showed biases against different religions, races, and genders.} (S01)
    \item \textit{Community members constantly showed racism and sexism.} (S01)
    \item \textit{Their financial inability to address an emotional crisis can undermine an already fragile condition, increasing feelings of hopelessness.} (S02)
    \item \textit{Users are asked to pay to access their own personal data.} (S02)
    \item \textit{One common experience was that features which were previously free were ‘taken away’.} (S02)
    \item \textit{After initially experiencing an app as amazing, the following reviewer felt betrayed and played after discovering that they could no longer engage in free gratitude exercises and goal setting.} (S02)
    \item \textit{Several users raised their concerns as apps did not offer the option to choose between male and female voices when interacting with bots.} (S03)
    \item \textit{Virtual assistant technologies are not as accurate or sensitive to the female voice as they are to the male voice.} (S03)
    \item \textit{Several comments raising requests to have more inclusive options on the choice of gender in account signup.} (S03)
    \item \textit{Users raised concerns about lack of inclusiveness and missing non-binary users.} (S03)
    \item \textit{Auto-correction feature of apps changes the gender of emojis chosen by users.} (S03)
    \item \textit{Users shared their frustrations with the content of the apps due to being sexist, gender violent, or being against their beliefs and value systems.} (S03)
    \item \textit{Users were concerned about companies promoting specific genders.} (S03)
    \item \textit{Companies making biased decisions when it comes to censoring content.} (S03)
    \item \textit{User feels that the developers deliberately make it difficult for the user to cancel their subscription.} (S09)
    \item \textit{Users complain of unfairness in either the process or outcome of the app.} (S09)
    \item \textit{User feels that they have not received a fair deal or that the app charges more money than it ought to.} (S09)
    \item \textit{The subscription model adopted by some apps resulted in them paying fees that were not commensurate with the services that they received.} (S17)
    \item \textit{Receiving different quality of features and services in different platforms and devices.} (S19)
    \item \textit{Feeling linguistic discrimination.} (S19)
    \item \textit{Exclude or ban some users based on their gender and race.} (S19)
    \item \textit{Biased censorship and promoting of a certain group of people.} (S19)
    \item \textit{Unfair, non-transparent, and sometimes misleading policies adopted by some apps for advertising and subscription.} (S19)
    \item \textit{It appears that they have little to no policies in place to protect their guests against a case manager who isn’t properly investigating their case.} (S23)
    \item \textit{Discriminated against for being superstraight.} (S29)
    \item \textit{They do not hire African Americans.} (S29)
    \item \textit{More fare than usual.} (S29)
    \item \textit{Suppress black lives matter, disabled, poor and anyone that’s not conventionally attractive.} (S29)
    \item \textit{It takes money and it doesn’t give so much robux.} (S31)
    \item \textit{Unfair payment pattern which makes subscription unaffordable or inconsiderate.} (S37)
\end{itemize}

\textbf{Accountability-related issues extracted by the primary studies from app reviews. }

\begin{itemize}
    \item \textit{Moderators have failed to follow the standard guidelines, resulting in frustrations among the users.} (S01)
    \item \textit{Red-flagged contents, such as “threatening suicide”, are left unchecked.} (S01)
    \item \textit{The apps frequently failed to provide support during emergency in a timely manner.} (S01)
    \item \textit{Long wait times and lack of follow-ups from their assigned coaches.} (S01)
    \item \textit{Received short, generic, out-of-context responses showing little professionalism from them.} (S01)
    \item \textit{Partial treatments also exacerbate MH conditions by increasing anxiety, especially for those who cannot afford upgrades.} (S02)
    \item \textit{Users expected better moderation from the app owners to remove these contents regularly, ban the accounts, or facilitate reporting such behaviours.} (S03)
    \item \textit{Users have informed the issues experienced while engaging customer support teams for complaint resolution.} (S08)
    \item \textit{Users perceive that the app provides false or inaccurate information.} (S09)
    \item \textit{User complains of not being able to access the app’s main functionality after purchase.} (S09)
    \item \textit{App developers are suspected of deleting reviews left by the user.} (S09)
    \item \textit{Lack of customer support and service/resolution center.} (S23)
    \item \textit{No policies in place to protect their guests against a case manager who isn’t properly investigating their case.} (S23)
    \item \textit{Creators aren’t held accountable for violating community guidelines even after users report them.} (S29)
    \item \textit{Pathetic customer service.} (S29)
    \item \textit{Couldn’t report for driver’s bad behavior.} (S29)
    \item \textit{Not banning cheaters.} (S31)
    \item \textit{Automated messages from customer support.} (S35)
    \item \textit{Difficulty in accessing developer assistance.} (S36)
    \item \textit{Therapeutic support found to be patronizing, impersonal, or inauthentic.} (S36)
    \item \textit{App developers or support staff are either unreachable, unresponsive, or unable to resolve (or slow in resolving) issues.} (S37)
\end{itemize}

\textbf{Safety-related issues extracted by the primary studies from app reviews. }

\begin{itemize}
    \item \textit{Users reported being subjected to abusive conduct and harassment.} (S01)
    \item \textit{Trolling, unpleasant behavior, and harassment were all common, along with negative comments about race, gender, and religion.} (S01)
    \item \textit{Blocking a user does not prevent them from harassing others.} (S01)
    \item \textit{Peers often encouraged unhealthy lifestyles and harmful thoughts, such as ``suicidal ideation”, ``self harm'', and ``dangerous self medicating''.} (S01)
    \item \textit{Communities on apps (e.g., social media, gaming groups) are violent or share hateful comments against specific genders.} (S03)
    \item \textit{Vulgar content (e.a., pornography, anti-society).} (S12)
    \item \textit{They were forced to use these tools that they perceived as creepy and dangerous.} (S14)
    \item \textit{Used for stalking and spying.} (S14)
    \item \textit{They don’t do any sort of background check.} (S23)
    \item \textit{Neither resolved my financial dispute, nor investigated this host or regarded my complaints/reports. I shouldn’t have had my hopes up about their safety or protection for guests or the credibility of their hosts} (S23)
    \item \textit{They allow dangerous medical misinformation and animal mistreatment to flourish.} (S29)
    \item \textit{Very dangerous tiktok ‘trends’ like spraying your bathroom mirror with a flammable substance and setting it on fire.} (S29)
    \item \textit{Spreading communal poison.} (S29)
    \item \textit{Advertising for gambling directed at kids.} (S29)
    \item \textit{Online harassment and cyber-bullying.} (S31)
    \item \textit{Age inappropriate and sexual content} (S32)
    \item \textit{Harmful advice and misdiagnosis.} (S36)
    \item \textit{Bullying and abuse.} (S36)
\end{itemize}

\section{Replication Packages} \label{apndxB}

\begin{itemize}
    \item \url{https://zenodo.org/records/8238376} (S03). This provides the annotated dataset of gender-related app reviews and automated methods for app review classification.
    \item \url{https://github.com/thekindlab/Dataset_Apps_Reviews_by_Accessibility_Guidelines} (S04). This provides a dataset of accessibility-related app reviews annotated with WCAG guidelines.
    \item \url{https://wajdialjedaani.github.io/A11yChi23/} (S05). This provides a dataset of app reviews associated with visual disabilities of eye conditions.
    \item \url{https://mhealthappsec.github.io/} (S06). This provides a dataset of app reviews related to privacy and security concerns in the health domain and the automated scripts used for mining and analysis of app reviews.
    \item \url{https://anonymous.4open.science/r/ml_app_reviews-3ED6/README.md} (S09). This provides a dataset of app reviews related to human values violations and scripts for collecting and classifying app reviews.
    \item \url{https://github.com/UBCFinder/UBCFinder} (S12). This provides a dataset of app reviews related to privacy and security, and automated scripts to identify undesired behaviors from app reviews.
    \item \url{https://github.com/marceloeler/data-ihc2019} (S15). This provides a dataset of accessibility-related app reviews and a comprehensive list of keywords used for extracting accessibility-related reviews.
    \item \url{https://zenodo.org/records/13139793} (S16). This provides a dataset of app reviews that specifically mention app updates and their impact on digital accessibility.
    \item \url{https://bit.ly/3e6wIiM} (S18). This provides a dataset of Portuguese app reviews related to accessibility in apps for autistic users.
    \item \url{https://zenodo.org/records/12180182} (S19). This provides a dataset of annotated transparency and fairness-related app reviews from AI-based mobile apps, and scripts for automated extraction and summarization of app reviews.
    \item \url{https://smilevo.github.io/access/CHI21_index.html} (S20). This provides a web service to automatically classify accessibility-related app reviews.
    \item \url{https://zenodo.org/records/5540624} (S24). This provides the scripts to perform automated sentiment analysis of accessibility-related app reviews.
    \item \url{https://zenodo.org/records/7755689} (S29). This provides a comprehensive dataset of manually annotated ethical concerns-related app reviews and scripts to perform automated classification of app reviews.
    \item \url{https://wajdialjedaani.github.io/MobileBB/} (S30). This provides a dataset of accessibility-related app reviews from the Blackboard app.
    \item \url{https://doi.org/10.6084/m9.figshare.25204388} (S32). This provides scripts to perform contextual analysis of ethical concerns-related app reviews.
    \item \url{https://seel.cse.lsu.edu/data/ase22.zip} (S34). This provides the dataset of privacy and security-related app reviews and scripts to perform summarization of app reviews.
\end{itemize}

\end{document}